\documentclass[a4paper,11pt]{elsarticle}

\usepackage[colorinlistoftodos]{todonotes}
\usepackage{placeins}
\usepackage{units}
\usepackage{siunitx}
\usepackage{graphicx} 
\usepackage{appendix}
\usepackage{amssymb}
\usepackage{slashed}
\usepackage[normalem]{ulem}
\usepackage{bbold}
\usepackage{comment}
\usepackage{xcolor}

\usepackage[T1]{fontenc} 
\usepackage{lineno,hyperref}
\usepackage{caption}
\usepackage{subcaption}

\begin{document}

\begin{frontmatter}

\title{\boldmath Wide-band full-wave electromagnetic modal analysis of the coupling between dark-matter axions and photons in microwave resonators}

\author[a]{P. Navarro}
\cortext[mycorrespondingauthor]{Corresponding author}
\ead{pablonm.ct.94@gmail.com}

\author[b]{Benito Gimeno*}
\ead{benito.gimeno@uv.es}

\author[a]{A. Álvarez Melcón}
\ead{alejandro.alvarez@upct.es}

\author[c]{S. Arguedas Cuendis}
\ead{sergio.arguedas.cuendis@cern.ch}

\author[d]{C. Cogollos}
\ead{cristian.cogollos@ub.edu}

\author[a]{A. Díaz-Morcillo}
\ead{alejandro.diaz@upct.es}

\author[e]{J.D. Gallego}
\ead{jd.gallego@oan.es}

\author[a]{J.M. García Barceló}
\ead{josemaria.gbarcelo@outlook.es}

\author[c,f]{J. Golm}
\ead{jessica.golm@cern.ch}

\author[g]{I.G. Irastorza}
\ead{Igor.Irastorza@cern.ch}

\author[a]{A.J. Lozano Guerrero}
\ead{antonio.lozano@upct.es}

\author[h,i]{C. Peña Garay}
\ead{cpenya@lsc-canfranc.es}

\address[a]{Department of Information and Communications Technologies, Technical University of Cartagena, 30203 - Cartagena, Spain}
\address[b]{Instituto de Física Corpuscular (IFIC), CSIC-University of Valencia, 46071 - Valencia, Spain}
\address[c]{European Organization for Nuclear Research (CERN), 1211 Geneva 23, Switzerland}
\address[d]{Instituto de Ciencias del Cosmos, University of Barcelona, 08028 - Barcelona, Spain}
\address[e]{Yebes Observatory, National Centre for Radioastronomy Technology and Geospace Applications,19080 - Guadalajara, Spain}
\address[f]{Institute for Optics and Quantum Electronics, Friedrich Schiller University Jena, Jena, Germany}
\address[g]{CAPA \& Departamento de Física Teórica, University de Zaragoza, 50009 - Zaragoza, Spain}
\address[h]{I2SysBio, CSIC-University of Valencia, 46071 - Valencia, Spain}
\address[i]{Laboratorio Subterráneo de Canfranc, 22880 - Estación de Canfranc, Huesca, Spain}

\begin{abstract}
{The electromagnetic coupling axion-photon in a microwave cavity is revisited with the Boundary Integral - Resonant Mode Expansion (BI-RME) 3D technique. Such full-wave modal technique has been applied for the rigorous analysis of the excitation of a microwave cavity with an axion field. In this scenario, the electromagnetic field generated by the axion-photon coupling can be assumed to be driven by equivalent electrical charge and current densities. These densities have been inserted in the general BI-RME 3D equations, which express the RF electromagnetic field existing within a cavity as an integral involving the Dyadic Green´s functions of the cavity (under Coulomb gauge) as well as such densities. This method is able to take into account any arbitrary spatial and temporal variation of both magnitude and phase of the axion field. Next, we have obtained a simple network driven by the axion current source, which represents the coupling between the axion field and the resonant modes of the cavity. With this approach, it is possible to calculate the extracted and dissipated RF power as a function of frequency along a broad band and without Cauchy-Lorentz approximations, obtaining the spectrum of the electromagnetic field generated in the cavity, and dealing with modes relatively close to the axion resonant mode. Moreover, with this technique we have a complete knowledge of the signal extracted from the cavity, not only in magnitude but also in phase. This can be an interesting issue for future analysis where the axion phase is an important parameter.}
\end{abstract}


\end{frontmatter}

\flushbottom

\section{Introduction}
The search of dark matter axions in the galactic halo has undergone an increasing activity in the last twenty years, following the experimental concept of a resonant haloscope from Sikivie \cite{Sikivie:1983ip}, \cite{Sikivie:2021} with the initial \cite{Asztalos:2001tf} (and still ongoing \cite{Braine:2019fqb}) work of ADMX, and continuing with other collaborations such as KLASH \cite{Alesini:2017ifp} at lower masses or HAYSTACK \cite{Kenany:2016tta}, ORGAN \cite{McAllister:2017lkb}, QUAX \cite{Alesini:2020vny}, CAPP \cite{Jeong:2017hqs},  or RADES \cite{RADES_paper1} at higher ones. A non-resonant dielectric haloscope (MADMAX) \cite{TheMADMAXWorkingGroup:2016hpc} has been proposed for even higher frequencies. All these experiments set the haloscopes realm currently in the 1 - 100 $\mu$eV axion mass range (240 MHz - 24 GHz in terms of frequency). A general review of these experiments and proper references can be found in \cite{Irastorza:2018dyq}.

Although the Lorentzian shape as an approximation of the resonant curve \cite{Peng:2000hd} is well-known in any mode supported by RF cavities, the expression of the peak value of the detected power \cite{Baker:2011na} is normally used for assessing the axion sensitivity of the proposed (or developed) experiment. In this paper we develop a complete semi-analytical solution for obtaining, through the BI-RME 3D method, the complex (magnitude and phase) current extracted from a low-loss cavity, where the axion-photon coupling due to the Primakoff effect takes place and, from it, the extracted RF power for a wide spectrum. This broadband rigorous result is specially interesting when neighboring modes are close to the axion one and can interfere with it. This situation can occur in large cavities with a high number of resonant modes or in multi-cavity haloscopes \cite{RADES_paper2}, where different configurations of the main mode can resonate at close proximity with the desired axion resonance. 

The paper is organized as follows. Section 2 provides a general view and detailed formulation of the BI-RME 3D method in resonant cavities. In section 3, the electromagnetic analysis of the axion-photon coupling under a static magnetic field is introduced, and this allows to apply the BI-RME 3D method to cavity haloscopes in section 4, specifically to two types of haloscopes: the cylindrical cavity decribed in  \cite{kim_CAPP_2019} and the rectangular multi-cavity reported in \cite{RADES_paper1}. Finally, section 5 summarizes the main conclusions of this work.

\section{The BI-RME 3D method}
The Boundary Integral Resonant Mode Expansion (BI-RME) method was developed during the eighties and nineties at the Universit\`{a} degli 
Studio di Pavia (Italy). It represents an advanced full-wave modal technique for the accurate and efficient electromagnetic analysis of microwave arbitrarily-shaped waveguides and cavities \cite{birme3d_3Dcavities}, \cite{birme3d_3Dcavities_ports} including metallic \cite{birme3d_fermin}, \cite{birme3d_angel_posts} and dielectric obstacles  \cite{birme3d_jordi_MWCL}, \cite{birme3d_jordi_MTT}, \cite{birme3d_pavia_MTT} of arbitrary geometry. The complete formulation and the different implementations are very extensive and can be found in the technical literature \cite{birme3d_overview}. 

\subsection{Lossless cavity}

Our starting point is to suppose that we have a microwave cavity resonator with arbitrary shape \cite{conciauro}. We will suppose that the volume of the cavity $V$ is simply connected. Inside the cavity we will assume that there is vacuum characterized by the electric permitivity $\varepsilon_0$ and the magnetic permeability $\mu_0$ of free space; dielectric and magnetic media can be accounted in the BI-RME theory, but they will not be considered in this work. 

\begin{figure}[h!]
	\begin{center}   
		\includegraphics[width=1 \textwidth]{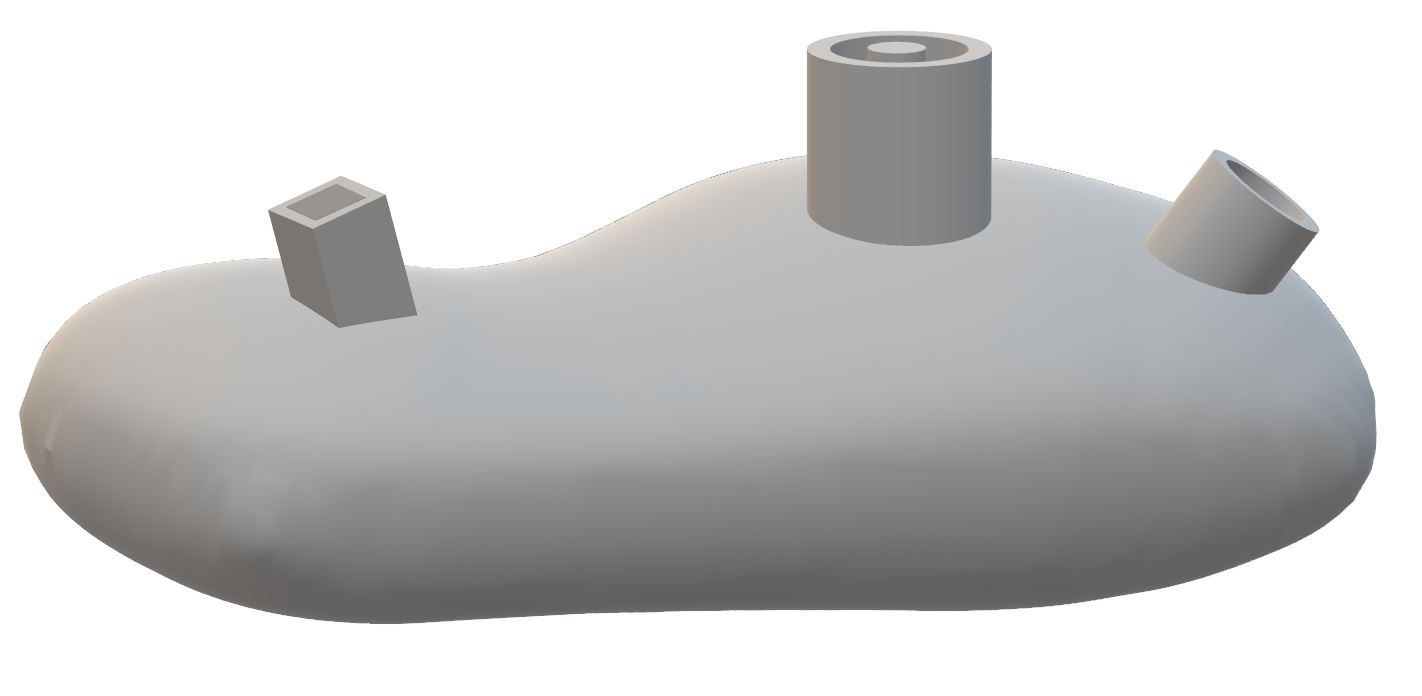}
		\caption{\label{fig:cavity_birme3d} Arbitrarily-shaped microwave resonant cavity connected to different access waveguide ports (rectangular, coaxial and circular).}
	\end{center}
\end{figure}

Let us consider a microwave resonant cavity with an arbitrary number of access waveguide ports $P$, as represented in Figure~\ref{fig:cavity_birme3d}. We will suppose that the conducting walls of the structure are lossless. The time-harmonic (phasors) electric and magnetic fields in such cavity originated by inner volumetric electric sources $\vec{J}$ and magnetic current sheets $\vec{M}$ can be expressed in terms of the electric and magnetic scalar and dyadic potentials (considering the Coulomb's gauge) as the following hybrid representation,
\begin{eqnarray}    
	\vec{E}(\vec{r}) & = & \frac{\eta}{j k} \nabla \int_V g^e
	(\vec{r},\vec{r}^{\, \prime}) \, \nabla' \cdot \vec{J}(\vec{r}^{\,
		\prime}) \, dV' - j k \eta \int_V \mathbf{\vec{G}^{\rm A}}
	(\vec{r},\vec{r}^{\, \prime}) \cdot \vec{J}(\vec{r}^{\, \prime})
	\, dV' - \nonumber \\[0.3cm] & & - \int_S \nabla \times \mathbf{\vec{G}^{\rm
			F}} (\vec{r},\vec{r}^{\, \prime}) \cdot \vec{M}(\vec{r}^{\,
		\prime}) \, dS' \, + \, \frac{1}{2} \, \vec{n} \times \vec{M}  \nonumber  
\end{eqnarray}	
\begin{eqnarray}  \label{birme3D_E_H_1}
	\vec{H}(\vec{r}) & = & \frac{1}{j k \eta} \nabla_s \int_S g^m
	(\vec{r},\vec{r}^{\, \prime}) \, \nabla' \cdot \vec{M}(\vec{r}^{\,
		\prime}) \, dS' - \frac{j k}{\eta} \int_S \mathbf{\vec{G}^{\rm F}}
	(\vec{r},\vec{r}^{\, \prime}) \cdot \vec{M} (\vec{r}^{\, \prime})
	\, dS' + \nonumber \\[0.25cm] & & + \int_V \nabla \times \mathbf{\vec{G}^{\rm A}}
	(\vec{r},\vec{r}^{\, \prime}) \cdot \vec{J}(\vec{r}^{\, \prime})   
	\, dV'
\end{eqnarray}
where $\eta = \sqrt{\mu_0/\varepsilon_0} \approx 377 \, \Omega$ is the vacuum impedance; $k=\omega/c$ is the free space wavenumber, $\omega$ being the angular frequency ($\omega = 2 \pi f$) and $c=1/\sqrt{\mu_0 \, \varepsilon_0}$ is the speed of light in vacuum; $j = \sqrt{-1}$ is the imaginary unit; $\vec{n}$ is the inward unitary normal vector to the cavity surface; $\nabla_s$ is the surface divergence operator \cite{chentotai}; $g^e(\vec{r},\vec{r}^{\, \prime})$ and $g^m(\vec{r},\vec{r}^{\, \prime})$ are the electric and magnetic static scalar potentials Green's functions of the cavity under Coulomb gauge, respectively; and $\mathbf{\vec{G}^{\rm A}}
(\vec{r},\vec{r}^{\, \prime})$ and $\mathbf{\vec{G}^{\rm F}} (\vec{r},\vec{r}^{\, \prime})$ are the electric and magnetic dyadic potentials Green's functions of the cavity under Coulomb gauge, respectively. In the Appendix we have summarized the most relevant properties of these Green's functions.

Next we insert the modal expansions of both electric and magnetic dyadic potential Green´s functions described in the Appendix in (\ref{birme3D_E_H_1}), obtaining
\begin{eqnarray}
	\vec{E}(\vec{r}) & = & \frac{- j \eta}{k} \, \nabla \int_V g^e
	(\vec{r},\vec{r}^{\, \prime}) \, \nabla' \cdot \vec{J}(\vec{r}^{\,
		\prime}) \, dV' - j k \eta \int_V \mathbf{\vec{G}^{\rm A}}_0
	(\vec{r},\vec{r}^{\, \prime}) \cdot \vec{J}(\vec{r}^{\, \prime})
	\, dV' - \nonumber \\[0.3cm] & & - \int_S \nabla \times \mathbf{\vec{G}^{\rm
			F}}_0 (\vec{r},\vec{r}^{\, \prime}) \cdot \vec{M}(\vec{r}^{\,
		\prime}) \, dS' \, + \, \frac{1}{2} \, \vec{n} \times \vec{M} \, +  \nonumber \\
	& & + \, (- j) k^3 \eta \, \sum_{m=1}^{+ \infty}  \frac{\vec{E}_m(\vec{r})}{k_m^2 (k_m^2 - k^2)} \, \int_V \vec{E}_m(\vec{r}^{\,\prime}) \cdot \vec{J}(\vec{r}^{\,\prime}) \, dV' \, + \nonumber \\ 
	& & + \,  (- k^2) \, \sum_{m=1}^{+ \infty}  \frac{\vec{E}_m(\vec{r})}{k_m (k_m^2 - k^2)} \, \int_S \vec{H}_m(\vec{r}^{\,\prime}) \cdot \vec{M}(\vec{r}^{\,\prime}) \, dS' \nonumber
\end{eqnarray}
\begin{eqnarray}   
\label{birme3D_E_H_2}
	\vec{H}(\vec{r}) & = & \frac{- j}{k \eta} \nabla_s \int_S g^m
	(\vec{r},\vec{r}^{\, \prime}) \, \nabla' \cdot \vec{M}(\vec{r}^{\,
		\prime}) \, dS' - \frac{j k}{\eta} \int_S \mathbf{\vec{G}^{\rm F}}_0
	(\vec{r},\vec{r}^{\, \prime}) \cdot \vec{M} (\vec{r}^{\, \prime})
	\, dS' + \nonumber \\[0.25cm] & & + \int_V \nabla \times \mathbf{\vec{G}^{\rm A}}_0
	(\vec{r},\vec{r}^{\, \prime}) \cdot \vec{J}(\vec{r}^{\, \prime})   
	\, dV' \, - \, \frac{j k^3}{\eta} \, \sum_{m=1}^{+ \infty}  \frac{\vec{H}_m(\vec{r})}{k_m^2 (k_m^2 - k^2)} \, \int_S \vec{H}_m(\vec{r}^{\,\prime}) \cdot \vec{M}(\vec{r}^{\,\prime}) \, dS' \, + \nonumber \\
	& & + \, k^2 \, \sum_{m=1}^{+ \infty}  \frac{\vec{H}_m(\vec{r})}{k_m (k_m^2 - k^2)} \, \int_V \vec{E}_m(\vec{r}^{\,\prime}) \cdot \vec{J}(\vec{r}^{\,\prime}) \, dV' 
\end{eqnarray}

Next, we need to model the excitation of the cavity using equivalent surface magnetic currents defined on the
access ports shown in Fig.\ref{fig:cavity_birme3d}. For this purpose, we first describe  the modes of the access waveguide ports. We will suppose that we have $P$ ports; in principle they might be different (rectangular, circular, coaxial, etc). The waveguide port $(\nu)$ is characterized by the electric and magnetic vector mode functions $\vec{e}^{(\nu)}_n$ and $\vec{h}^{(\nu)}_n$ \cite{marcuvitz} which satisfy the following relationships:
\begin{eqnarray}
	\int_{CS} \vec{e}_m^{(\mu)} \cdot \vec{e}_n^{(\nu)} \, dS \, & = & \, \int_{CS} \vec{h}_m^{(\mu)}
	\cdot \vec{h}_n^{(\nu)} \, dS \, = \, \delta_{m,n} \,  \delta_{\mu,\nu}     \nonumber \\
	\vec{n} \times \vec{e}_n^{(\nu)}   \, & = & \vec{h}_n^{(\nu)}  \, \, ;  \, \,   \mu, \nu = 1... P     \nonumber
\end{eqnarray}
This set of modes allows to express the transverse electromagnetic field in each waveguide port as a superposition of the vector mode functions:
\begin{eqnarray}
	\vec{E}_{transverse}^{(\nu)} \, = \, \sum_{n=1}^{+ \infty} V_n^{(\nu)} \, \vec{e}_n^{(\nu)}  \, \,  ;  \,  \,
	\vec{H}_{transverse}^{(\nu)} \, = \, \sum_{n=1}^{+ \infty} I_n^{(\nu)} \, \vec{h}_n^{(\nu)}  \nonumber 
\end{eqnarray}
where $V_n^{(\nu)}$ and $I_n^{(\nu)}$ are the voltage and current modal amplitudes of the $n$ mode at the port $(\nu)$, which are related by the modal impedance (admittance), $Z_n^{(\nu)}$ ($Y_n^{(\nu)}$), as follows:
\begin{eqnarray}
Z_n^{(\nu)} \, = \, \frac{1}{Y_n^{(\nu)}} \, = \, \frac{V_n^{(\nu)}}{I_n^{(\nu)}} 	  \nonumber
\end{eqnarray}
Thus the magnetic current sheets defined on the waveguide ports can be expressed in the form \cite{birme3d_3Dcavities_ports}:
\begin{eqnarray}    
\label{M_currents}
	\vec{M} \, = \, - \sum_{\nu = 1}^{P} \vec{n} \times \sum_{n=1}^{+ \infty} V_n^{(\nu)} \,  \vec{e}_n^{(\nu)} \, = \, -
	\sum_{\nu = 1}^{P} \, \sum_{n=1}^{+ \infty}  V_n^{(\nu)} \, \vec{h}_n^{(\nu)}
\end{eqnarray}

\begin{figure}[h!]
	\begin{center}   
		\includegraphics[width=1 \textwidth]{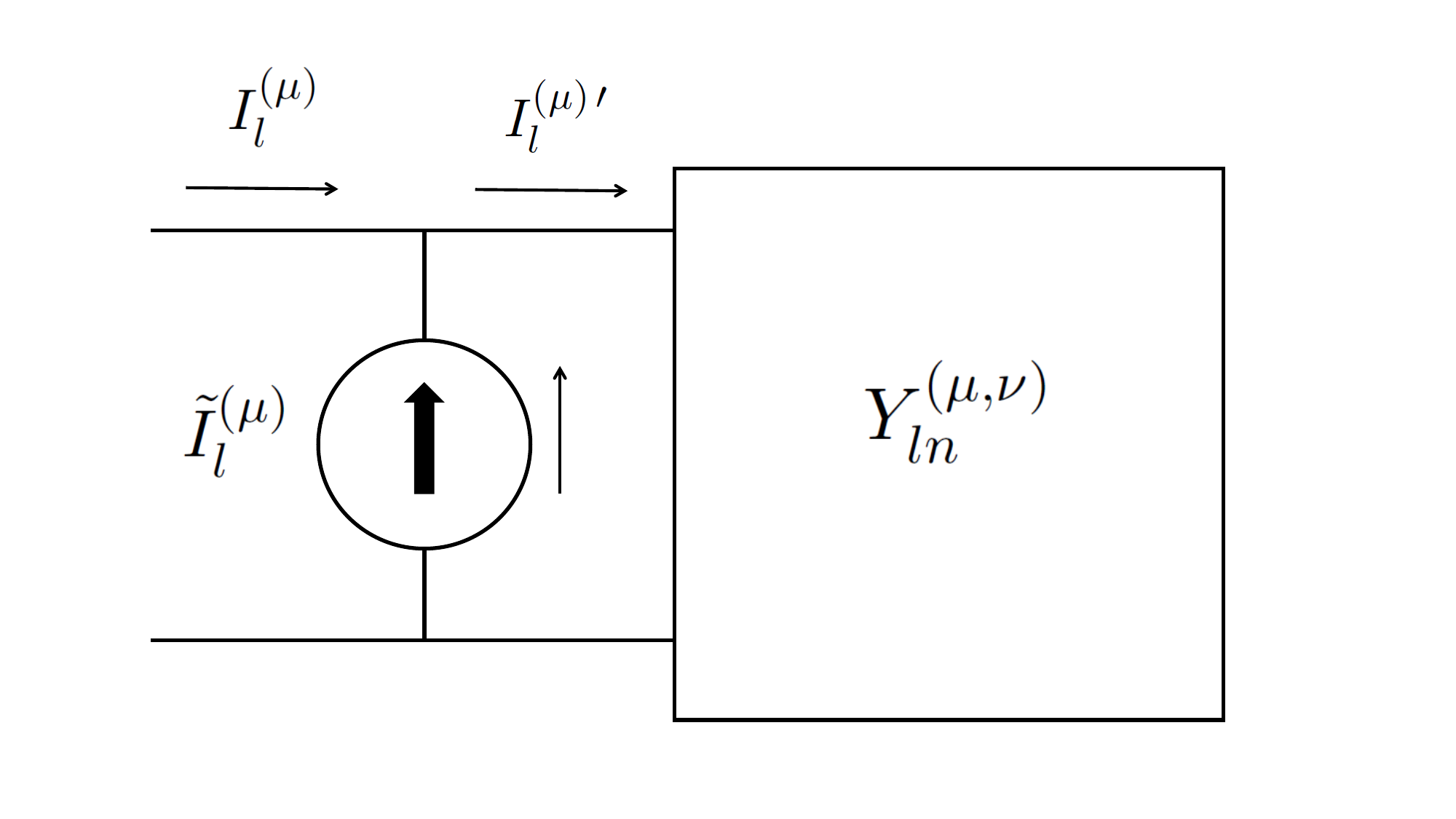}
		\caption{\label{fig:general_birme3d_network} Multimode equivalent network of a cavity resonator excited by an axion field. We have represented the port $(\mu)$.}
	\end{center}
\end{figure}

Next step in the BI-RME 3D formulation is to define the modal amplitudes $a_m$ as
\begin{eqnarray}
	a_m \, \equiv \, \frac{1}{k_m^2 (k_m^2 - k^2)} \, \left( j k \eta \, \int_V \vec{E}_m(\vec{r}^{\, \prime}) \cdot \vec{J}(\vec{r}^{\, \prime}) \, dV' \, - \, k_m  \sum_{\nu = 1}^{P} \, \sum_{n = 1}^{N_{\nu}} V_n^{(\nu)} \,  \int_{S(\nu)} \vec{H}_m(\vec{r}^{\, \prime}) \cdot \vec{h}_n^{(\nu)}(\vec{r}^{\, \prime}) \, dS' \right)   \nonumber
\end{eqnarray}
which are derived from (\ref{birme3D_E_H_2}) along with (\ref{M_currents}). After some algebraic manipulations, we find
\begin{eqnarray}
	\vec{E}(\vec{r}) & = & \frac{1}{2} \, \sum_{\nu = 1}^{P} \, \sum_{n = 1}^{+ \infty} V_n^{(\nu)} \, \vec{e}_n \, + \, 
	\frac{-j \eta}{k} \, \nabla \int_V g^e (\vec{r},\vec{r}^{\, \prime}) \, \nabla' \cdot \vec{J}(\vec{r}^{\,\prime}) \, dV' \, + \nonumber \\
	& &	+ \, (- j k \eta) \int_V \mathbf{\vec{G}^{\rm A}}_0(\vec{r},\vec{r}^{\, \prime}) \cdot \vec{J}(\vec{r}^{\, \prime})\, dV' \, - \, k^2 \, \sum_{m=1}^{+ \infty} \, a_m \, \vec{E}_m(\vec{r}) \, + \nonumber \\ & & 
	+ \, \sum_{\nu = 1}^{P} \, \sum_{n = 1}^{+ \infty} V_n^{(\nu)} \, \int_{S(\nu)} \nabla \times \mathbf{\vec{G}^{\rm
			F}}_0 (\vec{r},\vec{r}^{\, \prime}) \cdot \vec{h}_n^{(\nu)}(\vec{r}^{\,
		\prime}) \, dS'   \nonumber
\end{eqnarray}
\begin{eqnarray}
	\vec{H}(\vec{r}) & = & \frac{j}{k \eta} \, \sum_{\nu = 1}^{P} \, \sum_{n = 1}^{+ \infty} V_n^{(\nu)} \,  \nabla_s \int_{S(\nu)} g^m
	(\vec{r},\vec{r}^{\, \prime}) \, \nabla_s' \cdot \vec{h}_n^{(\nu)}(\vec{r}^{\,
		\prime}) \, dS' \, + \nonumber \\
	& &	 + \,  \frac{j k}{\eta} \,  \sum_{\nu = 1}^{P}  \, \sum_{n = 1}^{+ \infty} V_n^{(\nu)} \, \int_{S(\nu)} \mathbf{\vec{G}^{\rm F}}_0
	(\vec{r},\vec{r}^{\, \prime}) \cdot \vec{h}_n^{(\nu)} (\vec{r}^{\, \prime})
	\, dS' \,  + \, \int_ V \nabla \times \mathbf{\vec{G}^{\rm A}}_0
	(\vec{r},\vec{r}^{\, \prime}) \cdot \vec{J}(\vec{r}^{\, \prime})   
	\, dV' \, + \nonumber \\
	& & + \, \frac{- j k}{\eta} \, \sum_{m=1}^{+ \infty} \, a_m \, k_m \, \vec{H}_m(\vec{r})  \, + \, \frac{- j k}{\eta} \,  
	\sum_{\nu = 1}^{P} \, \sum_{n = 1}^{+ \infty} V_n^{(\nu)} \,  \sum_{m=1}^{+ \infty} \, \frac{\vec{H}_m(\vec{r})}{k_m^2} \, 
	\int_{S(\nu)} \, \vec{H}_m(\vec{r}^{\, \prime}) \cdot \vec{h}_n^{(\nu)}(\vec{r}^{\, \prime}) \, dS'
	\nonumber
\end{eqnarray}
Now we apply the boundary conditions: the tangential magnetic field on each waveguide port has to be a continuous function, because 
there are not any surface electric current defined at the port interface. As a consequence,   
\begin{eqnarray}
	\vec{H}(\vec{r}) |_{tangential \, on \,  S(\nu)} \, =  \, \sum_{\nu = 1}^{P} \, \sum_{n = 1}^{+ \infty} I_n^{(\nu)} \, \vec{h}_n^{(\nu)}(\vec{r})   
	\nonumber
\end{eqnarray}
Projecting both sides by the magnetic vector modal functions of the ports, and applying the orthonormalization condition among them, we find 
\begin{eqnarray}
	\int_{S(\nu)} \, \vec{H}(\vec{r}) |_{tang \, on \,  S(\nu)} \cdot \, \vec{h}_l^{(\mu)}(\vec{r}) \,  dS \, =  \, \int_{S(\nu)} \, \sum_{\nu = 1}^{P} \, \sum_{n = 1}^{+ \infty} I_n^{(\nu)} \, \vec{h}_n^{(\nu)}(\vec{r}) \cdot \, \vec{h}_l^{(\mu)}(\vec{r}) \,  dS \, = \,  I_l^{(\mu)}
	\nonumber
\end{eqnarray}
so the modal current amplitudes are expressed as
\begin{eqnarray}
	I_l^{(\mu)} \, & = & \, \int_{S(\mu)} \, \left( \frac{j}{k \eta} \, \sum_{\nu = 1}^{P} \, \sum_{n = 1}^{+ \infty} V_n^{(\nu)} \,  \nabla_s \int_{S(\nu)} g^m
	(\vec{r},\vec{r}^{\, \prime}) \, \nabla_s' \cdot \vec{h}_n^{(\nu)}(\vec{r}^{\,
		\prime}) \, dS' \right) \cdot \, \vec{h}_l^{(\mu)}(\vec{r}) \,  dS \,  + \nonumber \\
	& &	 + \,   \int_{S(\mu)} \, \left( \frac{j k}{\eta} \, \sum_{\nu = 1}^{P} \,   \sum_{n = 1}^{+ \infty} V_n^{(\nu)} \, \int_{S(\nu)} \mathbf{\vec{G}^{\rm F}}_0
	(\vec{r},\vec{r}^{\, \prime}) \cdot \vec{h}_n^{(\nu)} (\vec{r}^{\, \prime})
	\, dS' \, \right) \cdot \, \vec{h}_l^{(\mu)}(\vec{r}) \,  dS \, + \nonumber \\ 
	& &  + \, \int_{S(\mu)} \, \left( \int_ V \nabla \times \mathbf{\vec{G}^{\rm A}}_0
	(\vec{r},\vec{r}^{\, \prime}) \cdot \vec{J}(\vec{r}^{\, \prime})   
	\, dV' \,  \right) \cdot \, \vec{h}_l^{(\mu)}(\vec{r}) \,  dS  + \nonumber \\
	& & + \,  \int_{S(\mu)} \, \left( \frac{- j k}{\eta} \, \sum_{m=1}^{+ \infty} \, a_m \, k_m \, \vec{H}_m(\vec{r}) \right) \cdot \, \vec{h}_l^{(\mu)}(\vec{r}) \,  dS \, + \, \nonumber \\
	& &  \int_{S(\mu)} \, \left( \frac{- j k}{\eta} \,  
	\sum_{\nu = 1}^{P} \, \sum_{n = 1}^{+ \infty} V_n^{(\nu)} \,  \sum_{m=1}^{+ \infty} \, \frac{\vec{H}_m(\vec{r})}{k_m^2} \, 
	\int_{S(\nu)} \, \vec{H}_m(\vec{r}^{\, \prime}) \cdot \vec{h}_n^{(\nu)}(\vec{r}^{\, \prime}) \, dS' \right) \cdot \, \vec{h}_l^{(\mu)}(\vec{r}) \,  dS \nonumber
\end{eqnarray}
The surface divergence theorem (similar to the divergence theorem or Gauss theorem) \cite{chentotai}, \cite{hanson_yakovlev} has to be applied to the first integral, and after some algebraic manipulations we get
\begin{eqnarray}
	I_l^{(\mu)} \, & = & \, \frac{- j}{k \eta} \, \sum_{\nu = 1}^{P} \, \sum_{n = 1}^{+ \infty} V_n^{(\nu)} \, \int_{S(\mu)}  dS \,  \int_{S(\nu)}  dS' \,  \left( \nabla_s \cdot \, \vec{h}_l^{(\mu)}(\vec{r}) \right) \,  g^m
	(\vec{r},\vec{r}^{\, \prime}) \, \left( \nabla_s' \cdot \vec{h}_n^{(\nu)}(\vec{r}^{\,
		\prime}) \right) \,   + \nonumber \\
	& &	 + \,   \frac{j k}{\eta}   \, \sum_{\nu = 1}^{P}  \,  \sum_{n = 1}^{+ \infty} V_n^{(\nu)} \, \int_{S(\mu)} dS \int_{S(\nu)} dS' \,  \vec{h}_l^{(\mu)}(\vec{r}) \cdot \mathbf{\vec{G}^{\rm F}}_0
	(\vec{r},\vec{r}^{\, \prime}) \cdot \vec{h}_n^{(\nu)} (\vec{r}^{\, \prime})
	+ \nonumber \\ 
	& &  + \, \int_{S(\mu)} dS  \, \vec{h}_l^{(\mu)}(\vec{r}) \cdot \left( \int_ V \nabla \times \mathbf{\vec{G}^{\rm A}}_0
	(\vec{r},\vec{r}^{\, \prime}) \cdot \vec{J}(\vec{r}^{\, \prime})   
	\, dV' \,  \right)  \,   + \nonumber \\
	& & + \, \frac{- j k}{\eta} \, \sum_{m=1}^{+ \infty} \, a_m \, k_m \, \int_{S(\mu)} dS \,  \vec{h}_l^{(\mu)}(\vec{r}) \cdot \vec{H}_m(\vec{r}) \,   + \, \nonumber \\
	& &  + \,  \frac{- j k}{\eta} \,  
	\sum_{\nu = 1}^{P} \, \sum_{n = 1}^{+ \infty} V_n^{(\nu)} \,  \sum_{m=1}^{+ \infty} \, \frac{1}{k_m^2} \, \int_{S(\mu)} dS \, \vec{h}_l^{(\mu)}(\vec{r}) \cdot \vec{H}_m(\vec{r})
	\int_{S(\nu)} dS' \, \vec{h}_n^{(\nu)}(\vec{r}^{\, \prime}) \cdot \vec{H}_m(\vec{r}^{\, \prime}) \nonumber
\end{eqnarray}
In order to simplify this expression we define several matrices:
\begin{eqnarray}
	F_{mn}^{(\nu)} \, & \equiv & \,  \int_{S(\nu)} \vec{H}_m(\vec{r})  \cdot \vec{h}_n^{(\nu)}(\vec{r}) \, dS   \nonumber  \\
	G_{ln}^{(\mu,\nu)} \, & \equiv & \,   \int_{S(\mu)}  dS \,  \int_{S(\nu)}  dS' \,  \left( \nabla_s \cdot \, \vec{h}_l^{(\mu)}(\vec{r}) \right) \,  g^m
	(\vec{r},\vec{r}^{\, \prime}) \, \left( \nabla_s' \cdot \vec{h}_n^{(\nu)}(\vec{r}^{\,
		\prime}) \right) \,      \nonumber  \\
	T_{ln}^{(\mu,\nu)} \, & \equiv & \,   \int_{S(\mu)} dS \int_{S(\nu)} dS' \,  \vec{h}_l^{(\mu)}(\vec{r}) \cdot \mathbf{\vec{G}^{\rm F}}_0
	(\vec{r},\vec{r}^{\, \prime}) \cdot \vec{h}_n^{(\nu)} (\vec{r}^{\, \prime}) \nonumber
\end{eqnarray}
just obtaining
\begin{eqnarray}
	I_l^{(\mu)} \, & = & \, \frac{- j}{k \eta} \, \sum_{\nu = 1}^{P} \, \sum_{n = 1}^{+ \infty} V_n^{(\nu)} \, G_{ln}^{(\mu,\nu)}  \, + \,  
	\frac{j k}{\eta}   \, \sum_{\nu = 1}^{P}  \,  \sum_{n = 1}^{+ \infty} V_n^{(\nu)} \,  T_{ln}^{(\mu,\nu)} \, 
	+ \nonumber \\ 
	& &  + \, \int_{S(\mu)} dS  \, \vec{h}_l^{(\mu)}(\vec{r}) \cdot \left( \int_ V \nabla \times \mathbf{\vec{G}^{\rm A}}_0
	(\vec{r},\vec{r}^{\, \prime}) \cdot \vec{J}(\vec{r}^{\, \prime})   
	\, dV' \,  \right)  \,   + \nonumber \\
	& & + \, \frac{- j k}{\eta} \, \sum_{m=1}^{+ \infty} \, a_m \, k_m \,  F_{ml}^{(\mu)} \,   + \, 
	\frac{- j k}{\eta} \,  
	\sum_{\nu = 1}^{P} \, \sum_{n = 1}^{+ \infty} V_n^{(\nu)} \,  \sum_{m=1}^{+ \infty} \, \frac{1}{k_m^2} \, F_{ml}^{(\mu)} \, F_{mn}^{(\nu)} \nonumber
\end{eqnarray}
The integral containing the curl operator has to be properly treated using the properties of the dyadic Green functions; after inserting the modal coefficients $a_m$, we obtain
\begin{eqnarray}   
	I_l^{(\mu)} \, & = & \,  \sum_{\nu = 1}^{P} \, \sum_{n = 1}^{+ \infty} V_n^{(\nu)} \, \left( \frac{- j}{k \eta} \, G_{ln}^{(\mu,\nu)}  +   
	\frac{j k}{\eta}   \,   T_{ln}^{(\mu,\nu)}  +  \frac{j k^3}{\eta} \sum_{m=1}^{+ \infty} \, \frac{F_{ml}^{(\mu)} F_{mn}^{(\nu)}}{k_m^2(k_m^2 - k^2)} \right) \, +
	\nonumber \\ 
	& & + \, \sum_{m=1}^{+ \infty}  \, \frac{F_{ml}^{(\mu)}}{k_m} \,  \int_ V \vec{E}_m(\vec{r}^{\, \prime}) \cdot \vec{J}(\vec{r}^{\, \prime})   	\, dV'   + \, 
	k^2  \, \sum_{m=1}^{+ \infty} \,   \frac{F_{ml}^{(\mu)}}{k_m(k_m^2 - k^2)} \, \int_ V \vec{E}_m(\vec{r}^{\, \prime}) \cdot \vec{J}(\vec{r}^{\, \prime})   	\, dV' \,  .
	\nonumber
\end{eqnarray}
The first term is the BI-RME 3D expression of the generalized admittance matrix \cite{birme3d_3Dcavities_ports}, which is defined as
\begin{eqnarray}
	Y_{ln}^{(\mu,\nu)} \, \equiv \,   \frac{- j}{k \eta} \, G_{ln}^{(\mu,\nu)}  +   
	\frac{j k}{\eta}   \,   T_{ln}^{(\mu,\nu)}  +  \frac{j k^3}{\eta} \sum_{m=1}^{+ \infty} \, \frac{F_{ml}^{(\mu)} F_{mn}^{(\nu)}}{k_m^2(k_m^2 - k^2)} 
	\nonumber 
\end{eqnarray}
which completely characterizes the microwave device, and can be
viewed as an alternative representation to the generalized scattering matrix used by other authors \cite{collin_FMI}, \cite{pozar}. Thus the previous equation can be expressed in terms of the generalized admittance matrix as follows:
\begin{eqnarray}  \label{currents_non_truncated}
	I_l^{(\mu)} \, =  \,  \sum_{\nu = 1}^{P} \, \sum_{n = 1}^{+ \infty} \, Y_{ln}^{(\mu,\nu)} \,  V_n^{(\nu)} \,   +  \, 
	\sum_{m=1}^{+ \infty}  \, F_{ml}^{(\mu)} \, \frac{k_m}{k_m^2 - k^2}  \,  \int_ V \vec{E}_m(\vec{r}^{\, \prime}) \cdot \vec{J}(\vec{r}^{\, \prime})   	\, dV'    \,  .
\end{eqnarray}
This expression has been derived from the frequency domain Maxwell equations and it is exact. However, in a practical implementation of the algorithm both infinite series have to be truncated. The first infinite series related with the number of modes in each access port will be truncated to $N_{\nu}$ waveguide modes: typically we will have to include all the propagative modes of the waveguide and the first evanescent modes. For the second series, related to the cavity modes, we will have to include the first $M$ resonant modes existing in the analyzed frequency band. Thus, we will rewrite (\ref{currents_non_truncated}) as
\begin{eqnarray}   \label{currents_truncated}
	I_l^{(\mu)} \, = \,  \sum_{\nu = 1}^{P} \, \sum_{n = 1}^{N_{\nu}} \, Y_{ln}^{(\mu,\nu)} \,  V_n^{(\nu)} \,   +  \, 
	\sum_{m=1}^{M}  \, F_{ml}^{(\mu)} \, \frac{k_m}{k_m^2 - k^2}  \,  \int_ V \vec{E}_m(\vec{r}^{\, \prime}) \cdot \vec{J}(\vec{r}^{\, \prime})   	\, dV'    \,  .
\end{eqnarray}
Finally we define the following modal current sources:
\begin{eqnarray}   
\label{driven_source_birme3d}
	\tilde{I}_l^{(\mu)} \, & \equiv & \,  
	- \sum_{m=1}^M  \, F_{ml}^{(\mu)} \, \frac{k_m}{k_m^2 - k^2}  \,  \int_ V \vec{E}_m(\vec{r}^{\, \prime}) \cdot \vec{J}(\vec{r}^{\, \prime})   	\, dV'  
\end{eqnarray}	
and
\begin{eqnarray}
		I_l^{(\mu) \, \prime} \,  & \equiv &  \,  \sum_{\nu = 1}^{P} \, \sum_{n = 1}^{N_{\nu}} \, Y_{ln}^{(\mu,\nu)} \,  V_n^{(\nu)} \,  \nonumber
\end{eqnarray}
which allow to rewrite (\ref{currents_truncated}) in this form:
\begin{eqnarray}  \label{currents_extraction}
	I_l^{(\mu)} \,  =  \, I_l^{(\mu) \, \prime}  \,   -  \,  \tilde{I}_l^{(\mu)}  
\end{eqnarray}
A possible interpretation of this equation is shown in Figure~\ref{fig:general_birme3d_network}, and it can be easily understood in the context of classical Network Theory as reported in \cite{birme3d_angel_multipactor}: the set of current sources $\tilde{I}_l^{(\mu)}$ drives the cavity resonator characterized by its generalized admittance matrix $Y_{ln}^{(\mu,\nu)}$. It is worth to note that with the present formalism we have been able to describe the excitation of a cavity with a volumetric charge and current distribution, existing within the resonator due to the axion-photon coupling. Moreover, commercial software codes are not able to simulate the axion-photon excitation of a microwave passive component. For a better comprehension, we will apply the present formulation to two examples in section~4.   

\subsection{Cavity with lossy walls}
Considering a cavity with finite electrical conductivity $\sigma$, the effect of the Ohmic losses has to be accounted in the BI-RME 3D technique \cite{conciauro}. For such purpose we will use the perturbation method proposed in several books \cite{jackson}, \cite{collin_FMI}, \cite{collin_FTGW}, \cite{pozar}, \cite{kurokawa}. Thus, the lossless eigenvalues of the resonator $k_m$ have to be replaced by the perturbed eigenvalues of the lossy cavity $\kappa_m$:
\begin{eqnarray}
	k_m  \, \, \, \, \rightarrow \, \, \, \, \kappa_m  \, \approx  \,   k_m \, (1 - \frac{1}{2 \, Q_m}) \, + \, j \, \frac{k_m}{2 \, Q_m}
	\nonumber     
\end{eqnarray}
where $Q_m$ is the unloaded quality factor of the $m$ resonant mode defined as $Q_m \, =  \, \omega_m \, U_m/P_{c_{m}}$, $\omega_m = k_m \, c$ being the angular frequency of the $m$ mode, $U_m$ is the total time-average energy (electric and magnetic) stored in the cavity by the $m$ mode, and $P_{c_{m}}$ is the power loss of the $m$ mode, which is expressed as
\begin{eqnarray}
  P_{c_{m}} \, = \, \frac{R_s}{2} \, \int_{S_V} \, ||\vec{H}_m||^2 \, dS   \nonumber 
\end{eqnarray}
where $R_s = 1/(\sigma \delta)$ is the surface resistance of the conducting walls, and $\delta$ is the skin depth given by $\delta = \sqrt{2/(\omega \, \mu \, \sigma)}$ at room temperature, $\mu$ being the magnetic permeability of the conducting walls.

\section{Electromagnetic analysis of the interaction axion - photon}
Time-domain Maxwell's equations in SI units with the axion-photon interaction in the vacuum are given by
\begin{eqnarray}
	\nabla \cdot (\vec{\mathcal{E}} - c \, g_{a \gamma \gamma} \, \mathtt{a} \, \vec{\mathcal{B}}) \, & = & \, \frac{\mathcal{\varrho}_e}{\varepsilon_0}   \nonumber   \\
	\nabla \cdot \vec{\mathcal{B}} \, & = & \, 0  \nonumber  \\
	\nabla \times \vec{\mathcal{E}} \, & = & \, - \frac{\partial \, \vec{\mathcal{B}}}{\partial \, t}   \nonumber \\
	\nabla \times (c \, \vec{\mathcal{B}} + g_{a \gamma \gamma} \, \mathtt{a} \, \vec{\mathcal{E}}) \, & = & \, \frac{1}{c} \, \frac{\partial}{\partial \, t} (\vec{\mathcal{E}} - c \, g_{a \gamma \gamma} \, \mathtt{a} \, \vec{\mathcal{B}}) + c \, \mu_0 \, \vec{\mathcal{J}_e}   \nonumber
\end{eqnarray}
where $g_{a \gamma \gamma}$ is the two-photon coupling to an axion field constant. By assuming that the axion-photon interaction slightly modify the electromagnetic field, these equations can be decoupled into two parts \cite{kim_CAPP_2019}: one part for the external electromagnetic field $\vec{\mathcal{E}_e}$, $\vec{\mathcal{B}_e}$ generated by the classical charge $\mathcal{\varrho}_e$ and current $\vec{\mathcal{J}_e}$ densities, given by
\begin{eqnarray}  \label{maxwell_eq_td_classical}
	\nabla \cdot \vec{\mathcal{E}_e} \, & = & \, \frac{\mathcal{\varrho}_e}{\varepsilon_0}   \nonumber   \\
	\nabla \cdot \vec{\mathcal{B}_e} \, & = & \, 0            \nonumber  \\
	\nabla \times \vec{\mathcal{E}_e} \, & = & \, - \frac{\partial \, \vec{\mathcal{B}_e}}{\partial \, t}   \nonumber  \\
	\nabla \times \, \vec{\mathcal{B}_e} \, & = & \, \frac{1}{c^2} \, \frac{\partial \, \vec{\mathcal{E}_e}}{\partial \, t}  \,  +  \, \mu_0 \, \vec{\mathcal{J}_e}   
\end{eqnarray}
and another set of Maxwell's equations for the reacted fields $\vec{\mathcal{E}_a}$, $\vec{\mathcal{B}_a}$:  
\begin{eqnarray}  \label{maxwell_eq_td_axion}
	\nabla \cdot (\vec{\mathcal{E}_a} - c \, g_{a \gamma \gamma} \, \mathtt{a} \, \vec{\mathcal{B}_e}) \, & = & \,  0     \nonumber   \\
	\nabla \cdot \vec{\mathcal{B}_a} \, & = & \, 0   \nonumber  \\
	\nabla \times \vec{\mathcal{E}_a} \, & = & \, - \frac{\partial \, \vec{\mathcal{B}_a}}{\partial \, t}   \nonumber  \\
	\nabla \times (\vec{\mathcal{B}_a} + \frac{1}{c} \, g_{a \gamma \gamma} \, \mathtt{a} \, \vec{\mathcal{E}_e}) \, & = & \, \frac{1}{c^2} \, \frac{\partial}{\partial \, t} (\vec{\mathcal{E}_a} - c \, g_{a \gamma \gamma} \, \mathtt{a} \, \vec{\mathcal{B}_e})
\end{eqnarray}
The definition of the equivalent axion charge $\mathcal{\varrho}_a$ and current $\vec{\mathcal{J}_a}$ densities as 
\begin{eqnarray}  \label{axion_rho_J_densities}
	\mathcal{\varrho}_a  \, & \equiv & \,  g_{a \gamma \gamma} \, \sqrt{\frac{\varepsilon_0}{\mu_0}} \, \nabla \cdot (\mathtt{a} \, \vec{\mathcal{B}_e}) \nonumber   \\
	\vec{\mathcal{J}_a} \,  & \equiv & \, -  g_{a \gamma \gamma} \, \sqrt{\frac{\varepsilon_0}{\mu_0}} \,   
	\left( \frac{\partial \, (\mathtt{a} \, \vec{\mathcal{B}_e})}{\partial \, t} + \nabla \times (\mathtt{a} \, \vec{\mathcal{E}_e}) \right)   
\end{eqnarray}
allows to rewrite the axion Maxwell's equations (\ref{maxwell_eq_td_axion}) in the conventional form,
\begin{eqnarray}
	\nabla \cdot \vec{\mathcal{E}_a} \, & = & \, \frac{\mathcal{\varrho}_a}{\varepsilon_0}    \nonumber     \\
	\nabla \cdot \vec{\mathcal{B}_a} \, & = & \, 0   \nonumber   \\
	\nabla \times \vec{\mathcal{E}_a} \, & = & \, - \frac{\partial \, \vec{\mathcal{B}_a}}{\partial \, t}   \nonumber   \\
	\nabla \times \, \vec{\mathcal{B}_a} \, & = & \, \frac{1}{c^2} \, \frac{\partial \, \vec{\mathcal{E}_a}}{\partial \, t}  \,  +  \, \mu_0 \, \vec{\mathcal{J}_a}   \, .
\end{eqnarray}
Note that both set of charge and current densities satisfy the time-domain continuity equations:
\begin{eqnarray}
\nabla \cdot  \, \vec{\mathcal{J}_e} \, + \, \frac{\partial \, \mathcal{\varrho}_e}{\partial \, t}\,  =  \, 0 \, \, \,  \,  ; \, \,	\,  \,  \nabla \cdot  \, \vec{\mathcal{J}_a} \, + \, \frac{\partial \, \mathcal{\varrho}_a}{\partial \, t}\,  =  \, 0   \,  .  
\end{eqnarray}

Axion haloscopes are searching electromagnetic energy generated by the axion field within a microwave resonator in the presence of an external electric  $\vec{\mathcal{E}_e}$ and/or magnetic $\vec{\mathcal{B}_e}$ fields. In typical applications, only a very intense static magnetic field is applied, and the external electric field is zero $\vec{\mathcal{E}_e} = \vec{0}$. Thus, the field $\vec{\mathcal{B}_e}$ satisfies the classical Magnetostatic Maxwell's equations (\ref{maxwell_eq_td_classical}),
\begin{eqnarray}  
	\nabla \cdot \vec{\mathcal{B}_e} \, & = & \, 0            \nonumber  \\
	\nabla \times \, \vec{\mathcal{B}_e} \, & = & \,  \, \mu_0 \, \vec{\mathcal{J}_e}     \nonumber
\end{eqnarray}
where $\vec{\mathcal{J}_e} $ is the static external current density that creates the magnetostatic field $\vec{\mathcal{B}_e}$. As a consequence, the axion current density defined in (\ref{axion_rho_J_densities}) becomes
\begin{eqnarray}    \label{axion_current_B_static}	
	\vec{\mathcal{J}_a} \,  \equiv  \, -  g_{a \gamma \gamma} \, \sqrt{\frac{\varepsilon_0}{\mu_0}} \,   
	\frac{\partial \, (\mathtt{a} \, \vec{\mathcal{B}_e})}{\partial \, t}   \, = \,  -  g_{a \gamma \gamma} \, \sqrt{\frac{\varepsilon_0}{\mu_0}} \,  \vec{\mathcal{B}_e} \,   
	\frac{\partial \, \mathtt{a}}{\partial \, t} 
\end{eqnarray}

On the other hand, we will assume that the axion field is described by the axion electrodynamics equation (22) of the reference (\cite{Sikivie:2021}) which is expressed in natural units as 
\begin{eqnarray}    \label{equation_sikivie}
\frac{\partial^2 \mathtt{a}}{\partial t^2} \, - \, \nabla^2 \mathtt{a} \, + \, m_a^2 \, \mathtt{a} \, = \, - g_{a \gamma \gamma} \, \vec{\mathcal{E}} \cdot \vec{\mathcal{B}}.
\end{eqnarray}
where $m_a$ is the axion mass. By inserting the first-order expansion proposed in the subsection 2.2 of (\cite{kim_CAPP_2019}) for the electric $\vec{\mathcal{E}}	 \,  \approx \, \vec{\mathcal{E}}_e \, + \, g_{a \gamma \gamma} \, \vec{\mathcal{E}}_1 $ and the magnetic $\vec{\mathcal{B}}	 \,  \approx \, \vec{\mathcal{B}}_e \, + \, g_{a \gamma \gamma} \, \vec{\mathcal{B}}_1$ fields in (\ref{equation_sikivie}) and neglecting the terms $(g_{a \gamma \gamma})^2$ and higher-order terms, it is very easy to demonstrate the following result
\begin{eqnarray}
	\frac{\partial^2 \mathtt{a}}{\partial t^2} \, - \, \nabla^2 \mathtt{a} \, + \, m_a^2 \, \mathtt{a} = - g_{a \gamma \gamma} \, ( g_{a \gamma \gamma} \, \vec{\mathcal{E}}_1) \cdot (\vec{\mathcal{B}}_e \, + \, \vec{\mathcal{B}}_a)  = - (g_{a \gamma \gamma})^2 \, \vec{\mathcal{E}}_1 \cdot (\vec{\mathcal{B}}_e \, + \, \vec{\mathcal{B}}_a) \approx 0 \nonumber
\end{eqnarray}
which can be expressed in Fourier domain resulting as
\begin{eqnarray}
(\nabla^2 + \omega^2 - m_a^2) \, a \, = \, 0.  \nonumber
\end{eqnarray}
The dispersion relationship used by Sikivie  (\cite{Sikivie:2021})  is $\omega^2 \, = \, m_a^2 + k^2$ where $k$ is the magnitude of the axion wavenumber vector $\vec{k}$. Finally we obtain:
\begin{eqnarray}
	(\nabla^2 + k^2) \, a \, = \, 0. \nonumber
\end{eqnarray}
The complex phasor solution of this scalar Helmholtz wave equation can be expressed in terms of a plane wave,
\begin{eqnarray}    \label{axion_field_phasor}
	a(\vec{r}) \, = \, a_0 \, e^{-j (\vec{k} \cdot \vec{r} - \varphi)}   
\end{eqnarray}
where $a_0$ is the amplitude of the axion field and $\varphi$ is the initial phase. Thus, the real axion field can be easily calculated as follows,
\begin{eqnarray}
	\mathtt{a}(\vec{r},t) \, = \, \mathfrak{Re} (a \, e^{j \omega t}) \, = \,   \mathfrak{Re} (a_0  \, e^{j (\omega t -\vec{k} \cdot \vec{r} + \varphi)}) \, = \, a_0  \, \cos(\omega t -\vec{k} \cdot \vec{r} + \varphi)   \nonumber
\end{eqnarray}

At this point, the BI-RME 3D formalism can be used for the study of dark matter axions search haloscopes based on the concept of a microwave cavity with access waveguide ports. We will assume that the axion charge and current densities (\ref{axion_rho_J_densities}) are present in a lossy cavity. Then, equation (\ref{birme3D_E_H_1}) holds for such axion charge and current densities, so the BI-RME 3D can directly be applied. Following section~2, the source currents defined in (\ref{driven_source_birme3d}) become as
\begin{eqnarray}      \label{I_axion_full_notation_multimode}
	\tilde{I}_l^{(\mu)} \, & \equiv & \,  
	- \sum_{m=1}^M  \, F_{ml}^{(\mu)} \, \frac{k_m}{k_m^2 - k^2}  \,  \int_ V \vec{E}_m(\vec{r}^{\, \prime}) \cdot \vec{J_a}(\vec{r}^{\, \prime})   	\, dV'    
\end{eqnarray}
where $\vec{J_a}$ is the Fourier transform of the time-domain axion current density $\vec{\mathcal{J}_a}$. Next, we are going to apply the presented theory to the analysis of realistic haloscopes based on microwave resonators.

\section{Applications of the BI-RME 3D formalism to microwave haloscopes}
In this section the BI-RME 3D formalism is applied to resonant microwave haloscopes, and two practical examples are analyzed.

\subsection{BI-RME 3D formalism for resonant microwave haloscopes}

Next, we have to express the axion current density given in (\ref{axion_current_B_static}) in frequency-domain, obtaining
\begin{eqnarray}	\label{axion_J_phasor}
	\vec{J_a}(\vec{r}) \,  =  \, -  g_{a \gamma \gamma} \, \sqrt{\frac{\varepsilon_0}{\mu_0}} \,  \vec{\mathcal{B}_e}(\vec{r}) \, j \, \omega \,  a(\vec{r}) \, = \,   -  g_{a \gamma \gamma} \, \sqrt{\frac{\varepsilon_0}{\mu_0}} \,  \vec{\mathcal{B}_e}(\vec{r}) \, j \, \omega \, a_0 \, e^{j (- \vec{k} \cdot \vec{r} + \varphi)}
\end{eqnarray}

\begin{figure}[h!]
	\begin{center}   
		\includegraphics[width=1.2 \textwidth]{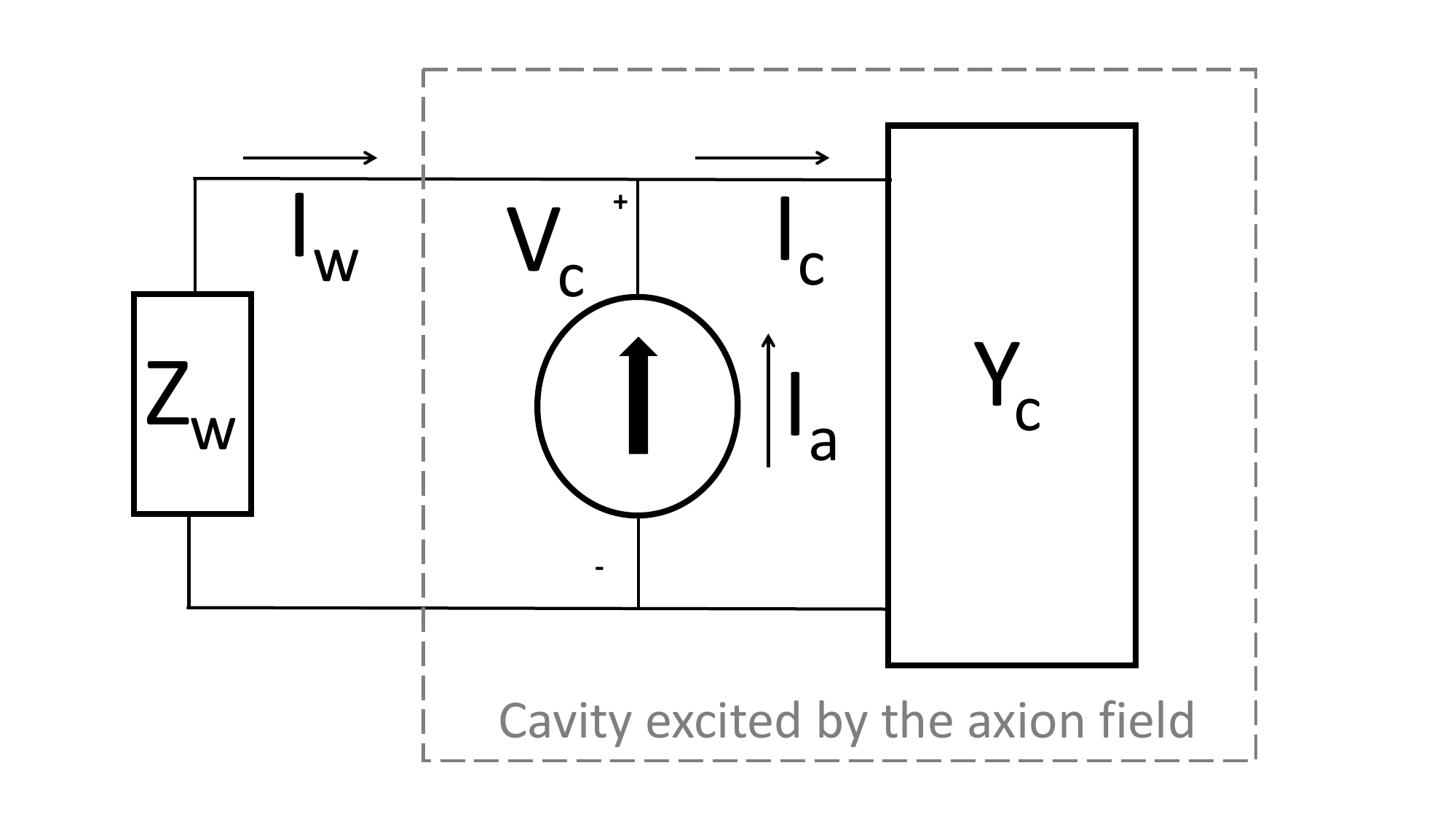}
		\caption{\label{fig:singleport_mode_birme3d_network} Single mode equivalent network of a cavity resonator excited by an axion field with one port.}
	\end{center}
\end{figure}

In order to simplify the implementation of the method, let us consider that we only have one port $P=1$ with one excited waveguide mode $N_1=1$. Thus, equation (\ref{currents_extraction}) becomes
\begin{eqnarray}    \label{current_1_full_notation}
	I_1^{(1)} \,  =  \,  Y_{11}^{(1,1)} \,  V_n^{(1)} \,   -  \,  \tilde{I}_1^{(1)}  \,  .
\end{eqnarray}
Now we will alleviate the notation by defining the following terms: the current extracted from the cavity $I_w \, \equiv \, I_1^{(1)}$, the cavity input admittance $Y_c \, \equiv \, Y_{11}^{(1,1)}$, the voltage in the cavity $V_c \, \equiv \,  V_1^{(1)}$, the current flowing on the cavity $I_c \, \equiv \, Y_c \, V_c$, and the current source generated by the axion-photon coupling $I_a \, \equiv \, \tilde{I}_1^{(1)}$.
This last current ($I_a$) is the unique generator in the network, so we can rewrite (\ref{current_1_full_notation}) as
\begin{eqnarray}  \label{currents_three}
	I_w \,  =  \,  Y_c \,  V_c \,   -  \,  I_a \, = \, I_c \, - \, I_a 
\end{eqnarray}
where we want to remark the definition of the axion current $I_a$ using equation (\ref{I_axion_full_notation_multimode}) and we have included the effect of the Ohmic losses:
\begin{eqnarray}     
\label{I_axion_one_port_M}
	I_a \, \equiv \,  \tilde{I}_1^{(1)} \, &  \equiv  & \,  
	- \sum_{m=1}^M \, F_{m1}^{(1)} \, \frac{\kappa_m}{\kappa_m^2 - k^2}  \,  \int_ V \vec{E}_m(\vec{r}^{\, \prime}) \cdot \vec{J_a}(\vec{r}^{\, \prime})   	\, dV' \, = \nonumber \\ & = &  \, \sum_{m=1}^{M}  \, \frac{\kappa_m}{k^2 - \kappa_m^2} \, \underbrace{\left( \int_{S(1)} \vec{H}_m(\vec{r})  \cdot \vec{h}_1^{(1)}(\vec{r}) \, dS \right)}_{COUPLING: CAV-PORT} \,   \,  \underbrace{\left( \int_ V \vec{E}_m(\vec{r}^{\, \prime}) \cdot \vec{J_a}(\vec{r}^{\, \prime})  \, dV'   \right)}_{COUPLING: AXION-CAV}
	\nonumber  \\    
\end{eqnarray}
This axion current is very important, since it represents the excitation of the cavity by the axion, and it relates both the axion-cavity coupling and the cavity-external port coupling. In Figure~\ref{fig:singleport_mode_birme3d_network} the single-mode equivalent network of the BI-RME 3D model based on (\ref{currents_three}) is presented: the axion current $I_a$ generated by the axion field (which acts as the current source) is divided into two branches, the current $I_c$ flowing along the cavity input admittance $Y_c$, and the current $- I_w$, flowing towards the external waveguide port, which is characterized by the modal impedance of the fundamental mode $Z_w$. Thus, we have demonstrated with a full-wave modal technique that the energy created by the axion within the cavity is split in two parts: the energy consumed by the cavity (Ohmic losses), and the energy extracted from the resonator towards the access waveguide port, which is the RF signal that we will be able to detect in an experimental test-bed. 

In this scenario we provide the expression of the cavity input admittance as a function of the reflection scattering parameter $S_{11}$ of the lossy cavity as \cite{pozar}
\begin{eqnarray}   \label{Yc_S11}
	Y_c \, = \, Y_w \,  \left( \frac{1 \, - \, S_{11}}{1 \, + \, S_{11}} \right)\, = \, G_c \, + \, j \, X_c
\end{eqnarray}
where $Y_w = 1/Z_w$ is the modal admittance of the fundamental waveguide mode of the port, and $G_c$ and $X_c$ are the real part and the imaginary part of $Y_c$, respectively.

Next, by applying the classical Network Theory we are able to calculate three time-average power terms: the power generated by the axion $P_a$, the power dissipated by the cavity $P_c$, and the power delivered to the the external access waveguide port $P_w$ as
\begin{eqnarray}
	V_c = \frac{I_a}{Y_w \, + \, Y_c}  = \frac{I_c}{Y_c} =  \frac{- I_w}{Y_w}  \,  \Rightarrow  \,   \left\{ \begin{array}{l} 	P_c  = \frac{1}{2}  Re(V_c \, I_c^*) =  \frac{|I_a|^2}{2 |Y _w \, + \, Y_c|^2} \, Re(Y_c^*)  \\
			P_w = \frac{1}{2} \, Re(V_c \, (- I_w)^*) = \frac{|I_a|^2}{2 |Y_w \, + \, Y_c|^2} \, Re(Y_w^*)
	\end{array} \nonumber
	\right.
	\nonumber
\end{eqnarray}
where the symbol $^*$ denotes conjugate complex.

Obviously, the Principle of Energy Conservation is satisfied: $P_a = P_c + P_w$ with $P_a  = (1/2)  Re(V_c \, I_a^*) = (1/2) |I_a|^2 Re(1/(Y_c+Y_w))$. At this point we want to emphasize that the cavity voltage $V_c$ is able to provide information of the measured RF signal phase, which is an important contribution of the BI-RME 3D formulation in comparison with other theories which are able to simulate only the detected power. It is very important to remark that in the present formulation we do not need to use any theoretical assumptions or approximations for the frequency spectrum dependency of the extracted power $P_w$, as the Cauchy-Lorentz \cite{kim_CAPP_2019} or Cauchy \cite{kim_CAPP_2020} distributions, used in previous contributions. What is more, the result is accurate even at frequencies far from the resonance peak.

It is worth noting that the presented theory can be applied to the calculation of the loaded quality factor of the $m$ resonant mode $Q_{L_m} = \omega_m U_m/(P_{w_m} \, + \, P_{c_m})$, where $P_{w_m}$ is the extracted power from the cavity to the access waveguide ports. Thus, the external quality factor of the $m$ resonant mode is defined as $Q_{e_m}  =  \omega_m \, U_m/P_{w_m}$. Therefore, the relationship among the three quality factors is $Q_{L_m}^{-1} = Q_{e_m}^{-1} + Q_m^{-1}$.

Now we define the cavity-waveguide coupling factor of the $m$ resonant mode $\beta_m$ as the ratio between the unloaded and external quality factors, resulting 
\begin{eqnarray}
	\beta_ m \, \equiv \, \frac{Q_m}{Q_{e_m}} \, = \, \frac{P_{w_m}}{P_{c_m}} \, = \, \frac{\frac{1}{2} \, \frac{|I_a|^2}{|Y_w \, + \, Y_c|^2} \, Re(Y_w^*)}{\frac{1}{2} \, \frac{|I_a|^2}{|Y_w \, + \, Y_c|^2} \, Re(Y_c^*)} \, = \, \frac{Re(Y_w^*)}{Re(Y_c^*)} = \, \frac{Y_w}{G_c}
	\nonumber
\end{eqnarray}
where we have assumed that the modal admittance $Y_w$ is real (the fundamental mode is propagative and Ohmic losses are neglected in the waveguide ports).

Microwave haloscopes typically operate under the critical coupling regime $\beta_m = 1$, which means that the power dissipated in the cavity (Ohmic losses) is equal to the power delivered to the waveguide ports, so $P_{c_m} = P_{w_m}$ or $Y_w = G_c$. By inserting this result in (\ref{Yc_S11}) it is very easy to demonstrate that the reflection coefficient of the cavity has to be zero ($S_{11} = 0$) under critical coupling regime, as it is well known. Furthermore, the overcoupled regime occurs when $\beta_m > 1$ which implies that $Y_w > G_c$, and in the undercoupled regime $\beta_m < 1$ so $Y_w < G_c$.

\subsection{Applications to microwave haloscopes with one coaxial waveguide port}
Now we are going to apply the present theory to cavities which have one coaxial access port. Such coaxial transmission line is characterized by the inner $b$ and the outer $a$ radii as well as the relative electrical permittivity of the dielectric medium between conductors $\varepsilon_r$. The normalized electric and magnetic vector mode functions of the fundamental TEM mode are given in cylindrical coordinates by \cite{marcuvitz}
\begin{eqnarray}
	\vec{e}_{TEM} \, = \, \frac{1}{\sqrt{2 \, \pi \, ln(a/b)}} \, \frac{1}{r} \, \hat{r}  \, \, \, \, ; \, \, \, \,  \vec{h}_{TEM} \, = \, \frac{1}{\sqrt{2 \, \pi \, ln(a/b)}} \, \frac{1}{r} \, \hat{\varphi}
	\nonumber
\end{eqnarray}
where the orthonormalization relationship used is
\begin{eqnarray}
	\int_S \, \vec{e}_{TEM} \cdot  (\vec{h}_{TEM} \times \hat{z}) \, dS \, = \, 1 \, .
	\nonumber
\end{eqnarray}
The modal impedance of the TEM coaxial mode is $Z_w = 1/Y_w = \sqrt{\mu_0/(\varepsilon_0 \varepsilon_r)}$, which should not be
confused with the characteristic coaxial impedance $Z_0 =  Z_w \, \ln(a/b)/(2 \pi)$.

In the analyzed cases, we will suppose that the Compton wavelength of the axions is high in comparison with the haloscope size, so the spatial dependence of the axion field can be neglected. As a consequence, equation (\ref{axion_field_phasor}) becomes $a \, = \, a_0 \, e^{j \,  \varphi}$ and  (\ref{axion_J_phasor}) results in
\begin{eqnarray}	
	\vec{J_a}(\vec{r}) \,  =  \,   -  g_{a \gamma \gamma} \, \sqrt{\frac{\varepsilon_0}{\mu_0}} \,  \vec{\mathcal{B}_e}(\vec{r}) \, j \, \omega \,  a_0 \, e^{j \varphi}  \, .
	\nonumber
\end{eqnarray}
Finally, this result must be inserted in (\ref{I_axion_one_port_M}) to obtain the axion current used in the simulations:
\begin{eqnarray}  \label{axion_current_coaxial_single_mode_port}
	I_a \,  = \, \frac{1}{\mu_0}  \, g_{a \gamma \gamma} \, a_0 \, e^{j \varphi} \, j \, k \,
	\sum_{m=1}^{M}  \, \frac{\kappa_m}{\kappa_m^2- k^2}   \, \left( \int_{S(1)} \vec{H}_m(\vec{r})  \cdot \vec{h}_{TEM}(\vec{r}) \, dS \right) \,   \,  \left( \int_ V \vec{E}_m(\vec{r}) \cdot  \vec{B}_e(\vec{r})  	\, dV   \right)  \nonumber  \\
\end{eqnarray}
with $\vec{B}_e \, \equiv \, \vec{\mathcal{B}_e}$ and $\vec{h}_{TEM} \equiv \vec{h}^{(1)}_1$. It is worth to note that the second integral is directly related with the geometric form factor $C_m$  \cite{kim_CAPP_2019}, which is defined in this scenario as
\begin{eqnarray}   \label{geometric_form_factor}
C_m \, \equiv \, \frac{\left|\int_ V \vec{E}_m(\vec{r}) \cdot  \vec{B}_e(\vec{r}) \, dV \right|^2}{\int_V \, ||\vec{B}_e(\vec{r})||^2 \, dV} 	
\end{eqnarray}
where simple bars  $| \cdot |$ indicate the magnitude of the complex number, and double bars  $|| \cdot  ||$ are the vector norm defined on the 3D complex vector space.

From an experimental point of view it is important to comment that the voltage measured in the coaxial waveguide port can be easily calculated as a function of the cavity voltage $V_c$ by integrating the electric field along the path from the inner to the outer radii,
\begin{eqnarray}
	V_{meas_a} \, - V_{meas_b} \, = \, - \int_b^a V_c \, \vec{e}_{TEM} \cdot d\vec{r}  \,   \, \Rightarrow \,  \, V_{meas} \, \equiv \, V_{meas_b} \, = \, V_c \, \sqrt{\frac{ln(a/b)}{2 \, \pi}}
	\nonumber
\end{eqnarray}
where we have assumed that the external conductor is the reference potential ($V_{meas_a} = 0$ V). Thus, $V_{meas}$ is the phasor (amplitude and phase) of the time-harmonic RF signal detected in the coaxial port.

Numerical simulations of this section have been computed with the commercial software CST STUDIO SUITE \cite{CST} and the postprocessing part has been developed with the commercial software MATLAB \cite{MATLAB}.

\subsubsection{Study of a cylindrical resonator cavity for benchmarking}
\begin{figure}[h!]
	\begin{center}   
		\includegraphics[width=1 \textwidth]{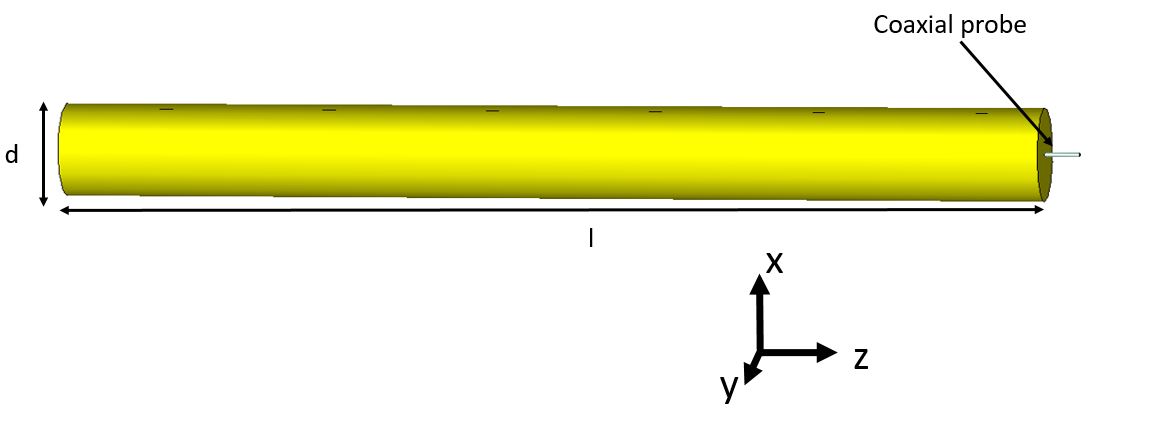}
		\caption{\label{fig:scheme_cylinder} Scheme of the cylindrical cavity.}
	\end{center}
\end{figure}

In order to compare this theory with the technical literature, we have benchmarked our algorithm with the example showed in Figure~4 of \cite{kim_CAPP_2019}. A cylindrical cavity with a diameter  $d = 90$~mm and a length $l = 1$~m has been used, as displayed in Figure~\ref{fig:scheme_cylinder}. The electrical conductivity of the metallic walls is $\sigma=6 \cdot  10^7$~S/m. A coaxial cable with characteristic impedance of $Z_0 = 50 \, \Omega$ ($b=0.635$~mm, $a=2.11$~mm, $\varepsilon_r=2.08$) has been inserted in the center of the top cap. The coaxial probe has been designed under critical coupling operation regime for the $TM_{010}$ mode ($f_ 1 \approx 2.55$~GHz). The axion field and the axion-photon interaction model is described by the parameters  $g_{a \gamma \gamma} \, a_0 = -8.51 \cdot 10^{-22}$ \cite{kim_CAPP_2019}. We have assumed an external homogeneous magnetostatic field oriented along the cylinder axis: $\vec{B}_e = B_e \, \hat{z}$ with $B_e=8$~T. The phase of the axion field used in the simulations is zero: $\varphi = 0$~rad. With these conditions, the axion current given in (\ref{axion_current_coaxial_single_mode_port}) becomes
\begin{eqnarray}    \label{I_axion_cylinder}
	I_a \,  = \, \frac{1}{\mu_0}  \, g_{a \gamma \gamma} \, a_0 \, B_e \, j \, k \,
	\sum_{m=1}^{M}  \, \frac{\kappa_m}{\kappa_m^2- k^2}   \, \left( \int_{S(1)} \vec{H}_m(\vec{r})  \cdot \vec{h}_{TEM}(\vec{r}) \, dS \right) \,   \,  \left( \int_ V \vec{E}_m(\vec{r}) \cdot  \hat{z}  	\, dV   \right)   \nonumber  \\
\end{eqnarray}

\begin{figure}[h!]
	\centering
	    \begin{subfigure}[b]{0.8\textwidth}
	        \includegraphics[width= \textwidth]{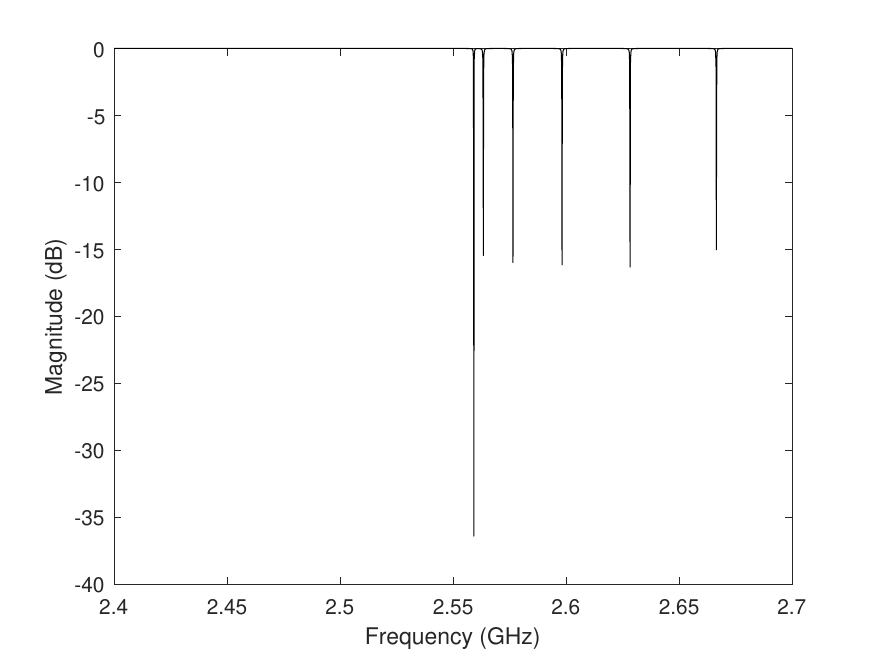}
	    \end{subfigure}
	   		\hfill
        \begin{subfigure}[b]{0.8\textwidth}
	            \includegraphics[width= \textwidth]{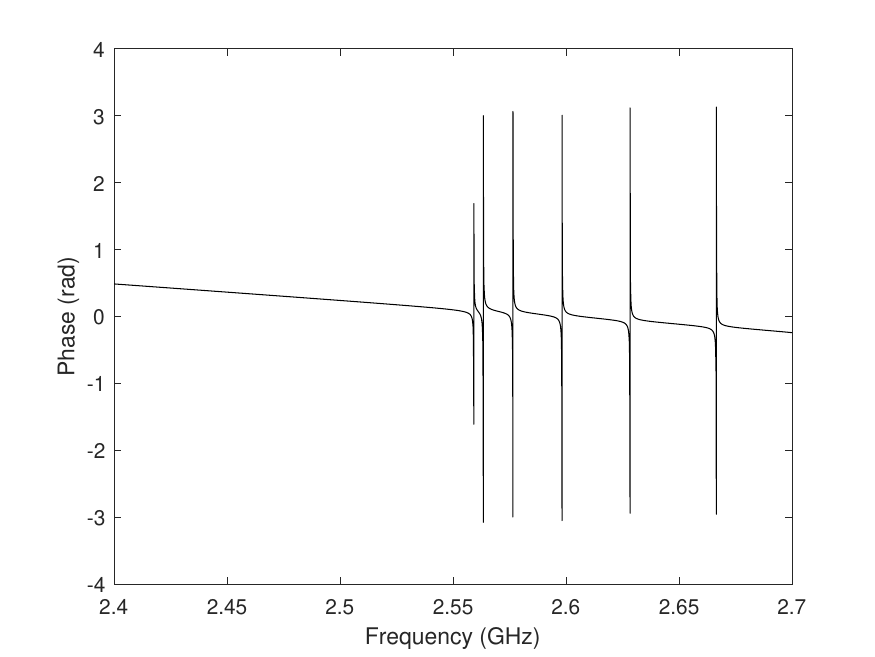}
	    \end{subfigure}		
		\hfill
		\caption{\label{fig:S11_cylinder} Reflection scattering parameter $S_{11}$ as a function of the frequency for the cylindrical resonator. Magnitude (top) and phase (bottom) have been plotted.}
\end{figure}

In Figure~\ref{fig:S11_cylinder} both magnitude and phase of the reflection scattering coefficient $S_{11}$ of the cylindrical cavity obtained with CST Studio are plotted. We can observe the resonances of the modes excited by the coaxial probe. In Table~I the most relevant parameters of the structure for these set of modes are shown: resonant frequencies $f_m$, unlodaded quality factors $Q_m$ and geometric form factors $C_m$ as defined in (\ref{geometric_form_factor}). We can see that the geometric form factor of the first mode is significantly higher than the others; in particular, the factor of the last three modes is negligible.
\begin{table}
\begin{center}
\caption{Parameters of the cylindrical cavity.}
\begin{tabular}{|c|c|c|c|c|}    \hline
    $m$ & Mode & $f_m$ (GHz) &  $Q_m$	& $C_m$  \\  \hline
  1     & $TM_{010}$  & $2.55920$  &  $33069.1$  &  $0.68505$  \\  \hline
  2     & $TM_{011}$  & $2.56334$  &  $33260.6$  &  $0.00523$  \\  \hline  
  3     & $TM_{012}$  & $2.57646$  &  $33119.7$  &  $0.00027$  \\  \hline  
  4     & $TM_{013}$  & $2.59817$  &  $33292.4$  &  $5.305 \cdot 10^{-5}$  \\  \hline 
  5     & $TM_{014}$  & $2.62827$  &  $33493.8$  &  $1.710 \cdot 10^{-5}$  \\  \hline  
  6     & $TM_{015}$  & $2.66646$  &  $33781.9$  &  $7.247 \cdot 10^{-6}$  \\  \hline  
\end{tabular}
\end{center}
\end{table}

\begin{figure}[h!]
	\begin{center}   
		\begin{subfigure}[b]{0.8\textwidth}
	            \includegraphics[width= \textwidth]{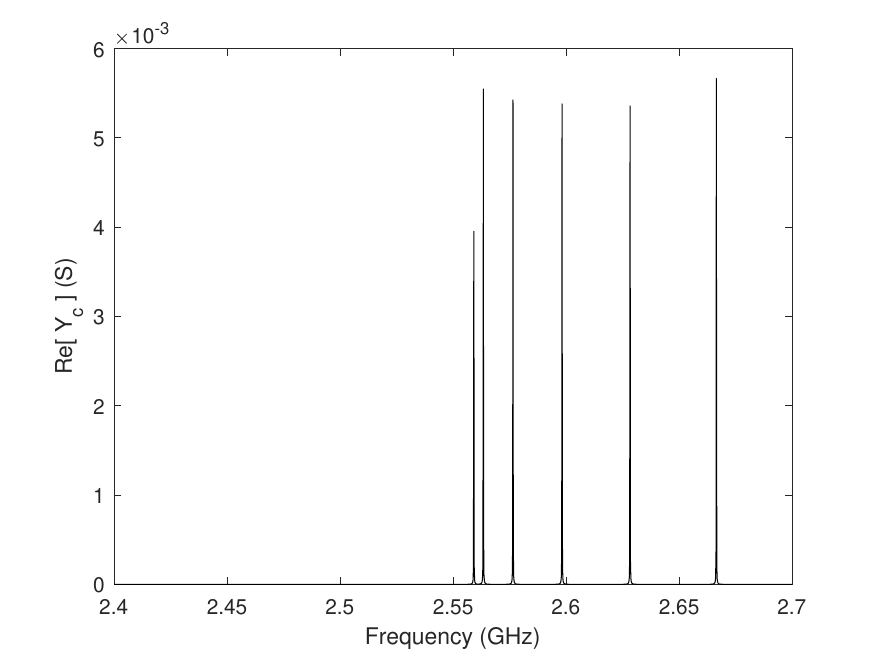}
	    \end{subfigure}
        \begin{subfigure} [b]{0.8\textwidth}
	            \includegraphics[width= \textwidth]{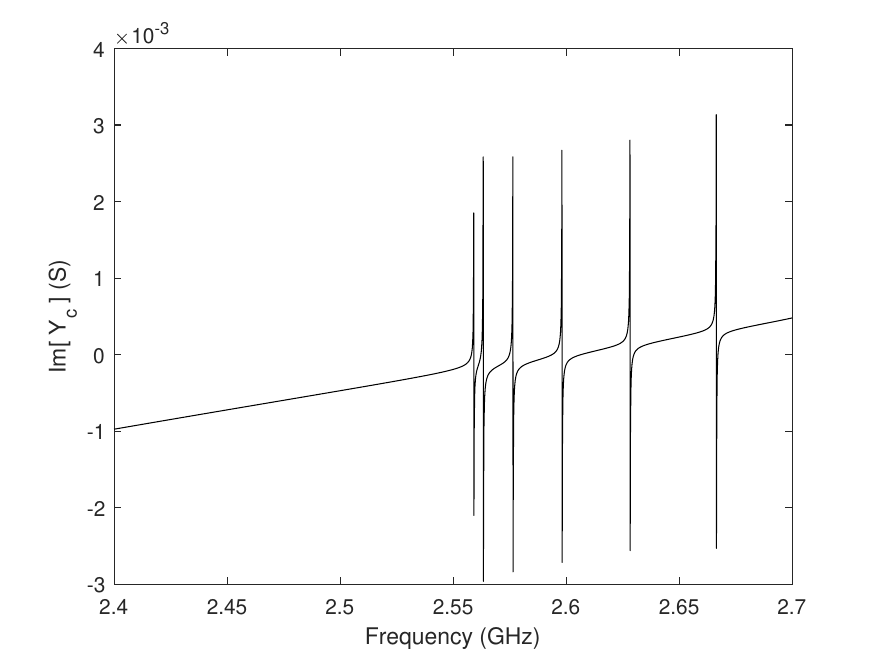}
	    \end{subfigure}	
		\caption{\label{fig:Yc_cylinder} Input admittance of the cavity $Y_c$ as a function of the frequency for the cylindrical resonator. Real part (top) and imaginary part (bottom) are shown.}
	\end{center}
\end{figure}

Next, the input admittance of the cavity $Y_c$ computed with equation (\ref{Yc_S11}) is presented in Figure~\ref{fig:Yc_cylinder}. Both the real $G_c$ and the imaginary $X_c$ parts are plotted as a function of frequency. We want to point out that the real part is always positive because it represents a passive resistive behaviour, and the imaginary part is zero at the resonant frequencies, according with the classical Network Theory. Moreover, the positive sign of the imaginary part denotes a capacitive behaviour, whereas if the sign is negative it represents an inductive one.  

\begin{figure}[h!]
	\begin{center}   
	    \begin{subfigure} [b]{0.8\textwidth}
	            \includegraphics[width= \textwidth]{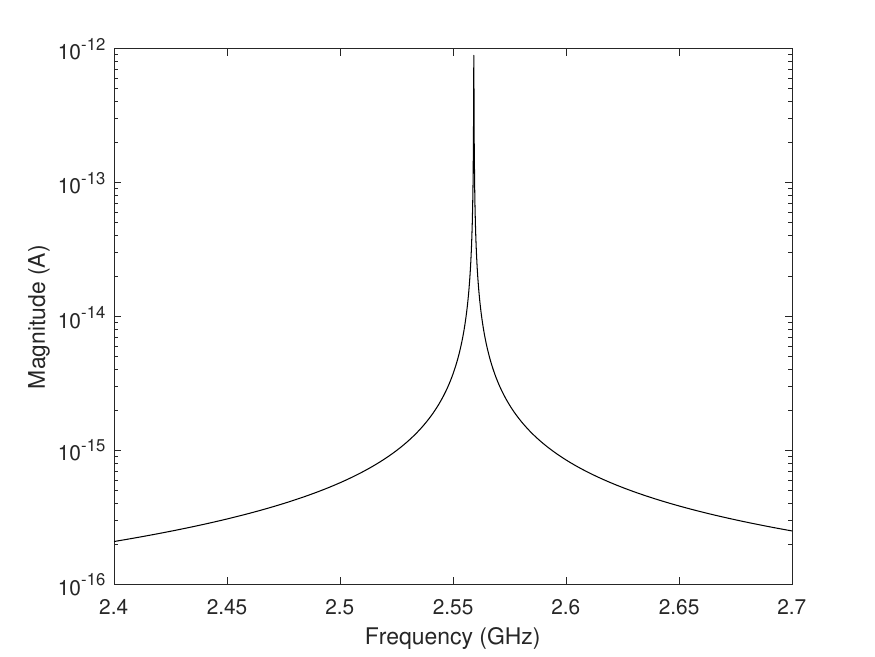}
	    \end{subfigure}
        \begin{subfigure} [b]{0.8\textwidth}
	            \includegraphics[width= \textwidth]{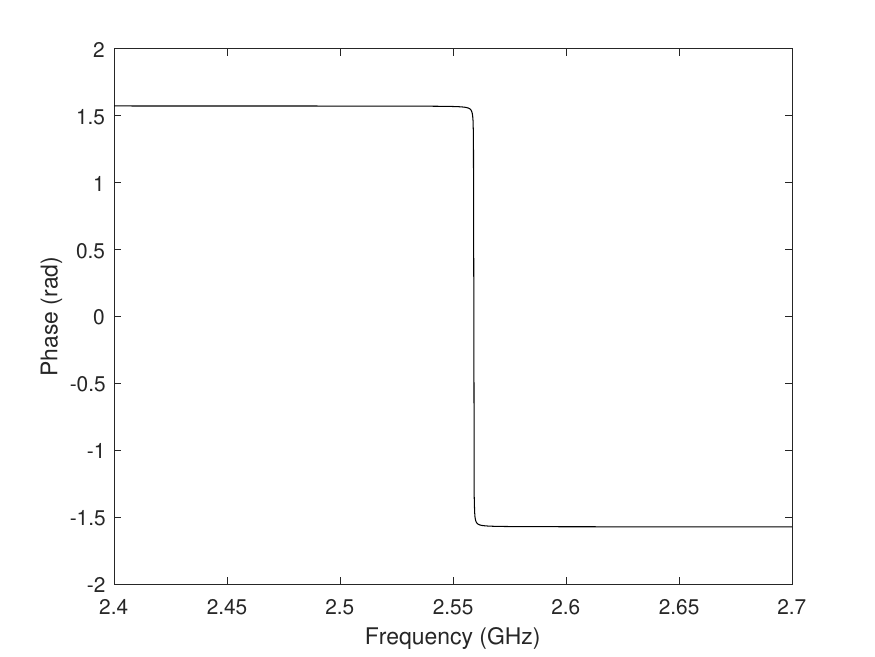}
	    \end{subfigure}	
		\caption{\label{fig:Ia_M1_cylinder} Axion current $I_a$ of the cylindrical cavity as a function of the frequency considering only the first mode ($M=1$). Magnitude (top) and phase (bottom) are plotted.}
	\end{center}
\end{figure}

\begin{figure}[h!]
	\begin{center}   
	     \begin{subfigure} [b]{0.8\textwidth}
	            \includegraphics[width= \textwidth]{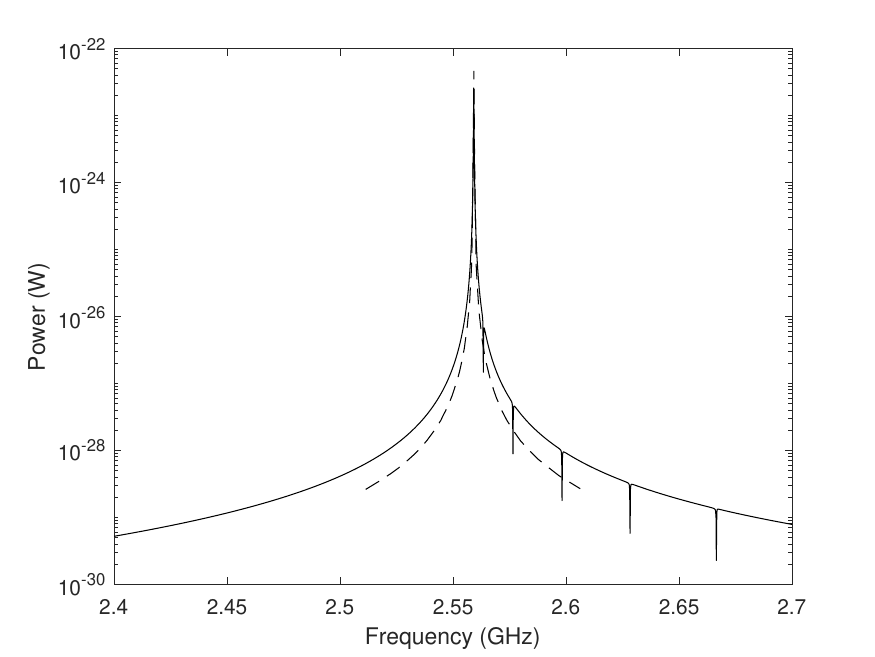}
	    \end{subfigure}
        \begin{subfigure} [b]{0.8\textwidth}
	            \includegraphics[width= \textwidth]{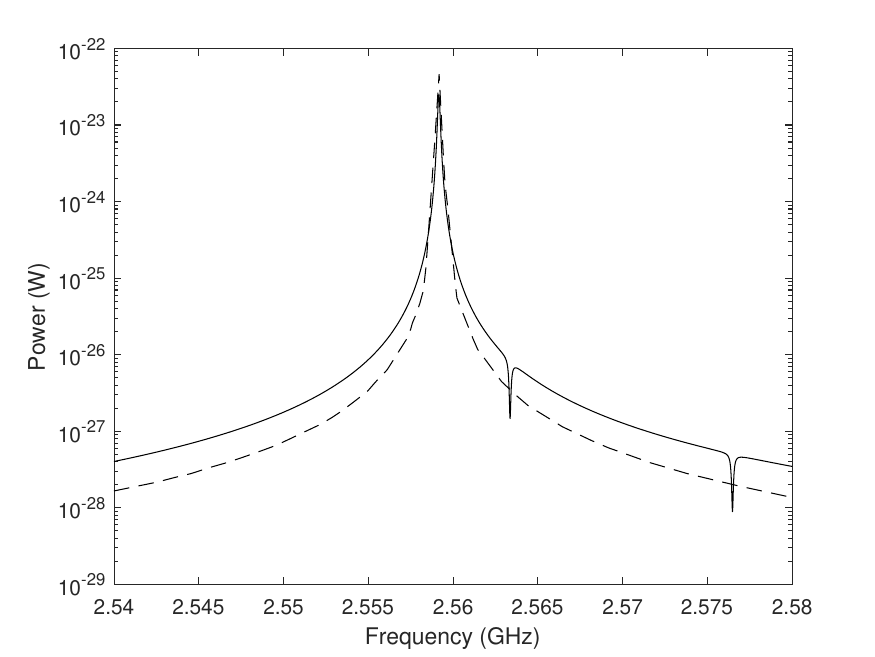}
	    \end{subfigure}	
		\caption{\label{fig:power_M1_cylinder} Extracted power $P_w$ from the cylindrical cavity (continuous line) as a function of the frequency considering only the first mode ($M=1$) in comparison with \cite{kim_CAPP_2019} (dashed line) (top). We also show a zoom in the bottom figure for a closer comparison.}
	\end{center}
\end{figure}

In order to compare our technique with \cite{kim_CAPP_2019}, we have computed the axion current with equation (\ref{I_axion_cylinder}) including only the first resonant mode ($M=1$). In Figure~\ref{fig:Ia_M1_cylinder} we plot $I_a$ in both magnitude and phase as a function of frequency, showing the peak of the resonant mode $TM_{010}$ according to Table~I. In Figure~\ref{fig:power_M1_cylinder} the delivered power to the coaxial port ($P_w$) computed with both methods is compared, demonstrating a good agreement not only in the resonant frequency but also in a wide frequency range, which allows to validate the presented procedure. 

\begin{figure}[h!]
	\begin{center}
	    \begin{subfigure} [b]{0.8\textwidth}
	            \includegraphics[width= \textwidth]{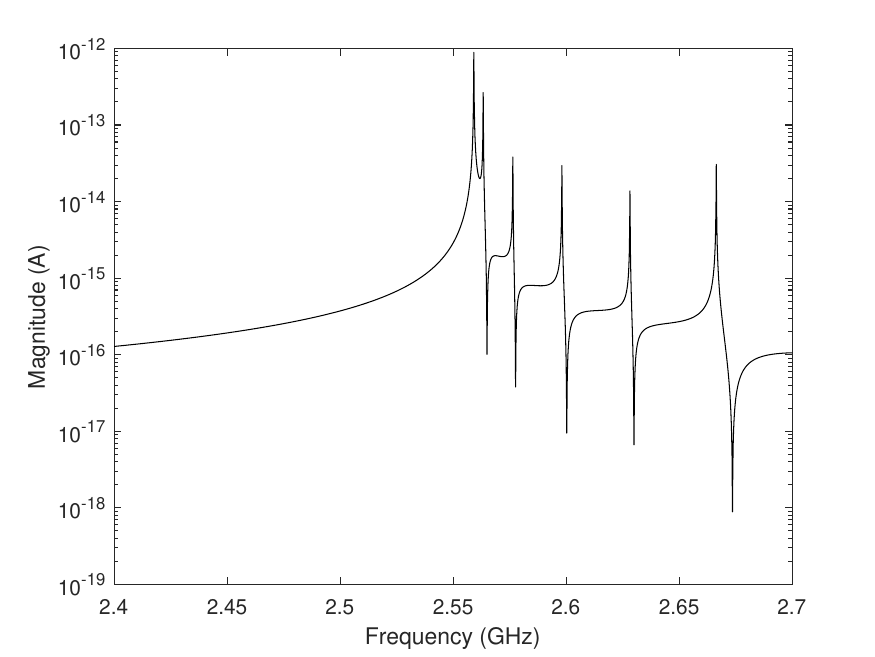}
	    \end{subfigure}
        \begin{subfigure} [b]{0.8\textwidth}
	            \includegraphics[width= \textwidth]{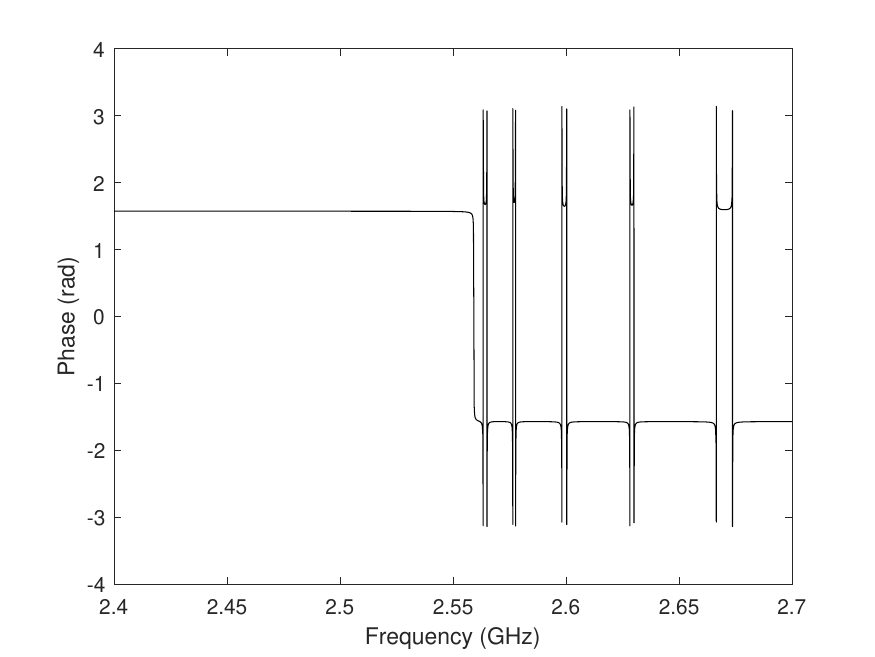}
	    \end{subfigure}	
		\caption{\label{fig:Ia_M6_cylinder} Axion current $I_a$ of the cylindrical cavity as a function of the frequency, considering the total set of modes coupled with the coaxial probe ($M=6$). Magnitude (top) and phase (bottom) are displayed.}
	\end{center}
\end{figure}

\begin{figure}[h!]
	\begin{center}   
        \begin{subfigure} [b]{0.8\textwidth}
	        \includegraphics[width= \textwidth]{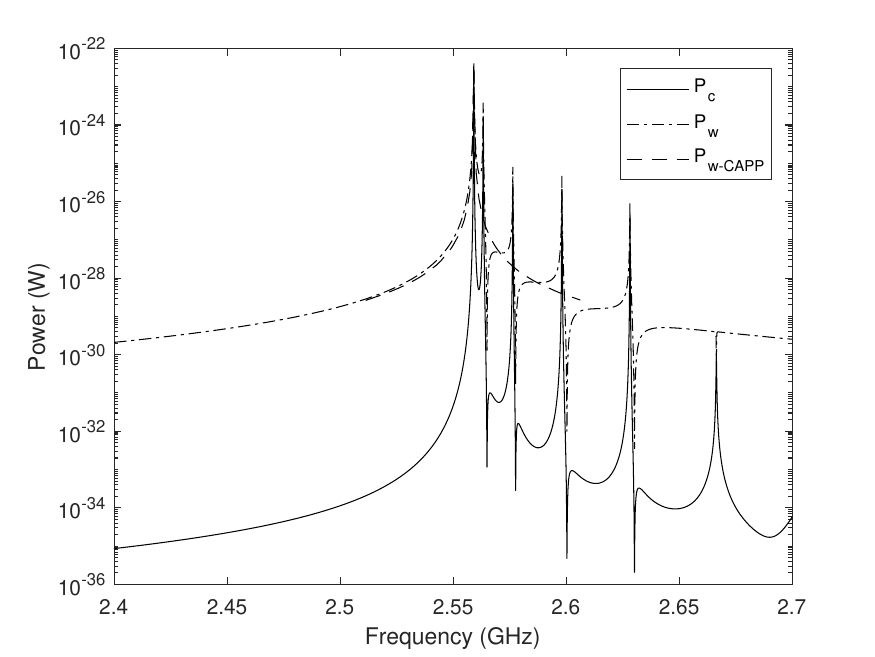}
	    \end{subfigure}
        \begin{subfigure} [b]{0.8\textwidth}
	        \includegraphics[width= \textwidth]{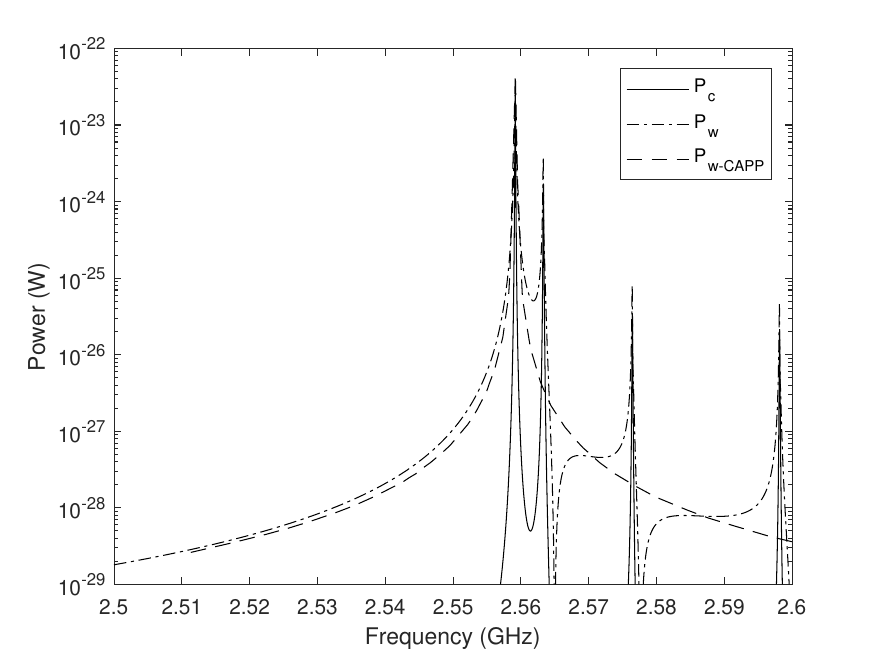}
	    \end{subfigure}			
	    \caption{\label{fig:power_M6_cylinder} Extracted $P_w$ and dissipated $P_c$ powers of the cylindrical cavity as a function of the frequency, considering the full set of modes coupled with the coaxial probe ($M=6$) in comparison with \cite{jackson} (top). In the bottom figure, we present a zoom of the plot in order to check the coupling regime achieved in the design of the coaxial probe, observing a very accurate critical coupling condition.}
	\end{center}
\end{figure}

Next, we have computed the electrical response including the full set of modes coupled with the coaxial monopole ($M=6$) in the analyzed frequency range, as reported in Table~I. In Figure~\ref{fig:Ia_M6_cylinder} the axion current (\ref{I_axion_cylinder}) is depicted, observing the contribution of each resonant mode to the equivalent axion current source. It is evident that the greater the modal geometric form factor (\ref{geometric_form_factor}), the higher the axion current is. Both, the power delivered to the coaxial port $P_w $ as well as the power dissipated in the cavity $P_c$ are showed in Figure~\ref{fig:power_M6_cylinder}. In this plot we have also included the simulations presented in \cite{kim_CAPP_2019}. In order to provide an accurate result, it is remarkable that we have to consider the complete set of excited modes in the analyzed frequency range, which precisely represents the global modal spectrum of the cavity. In this sense, our simulation is more precise than the method presented in \cite{kim_CAPP_2019} because it is able to account for the contribution of all the resonant modes excited within the resonator. Therefore, we can state that the BI-RME 3D is a wide-band numerical technique, since it allows to obtain the frequency response of the system in the explored frequency range. Finally, we observe the frequency response of the dissipated power in the cavity, demonstrating that at the resonant peak of the fundamental mode $TM_{010}$ the power lost in the cavity is equal to the extracted power, as predicted in the critical coupling regime.

\subsubsection{Study of the first RADES all-inductive coupled cavities structure}
\begin{figure}[h!]
	\begin{center}   
		\includegraphics[width=0.8 \textwidth]{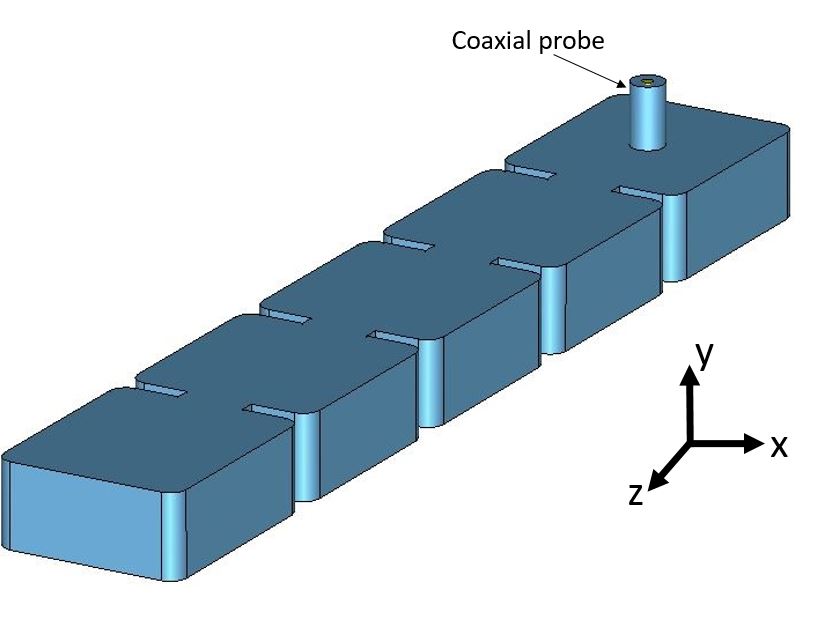}
		\caption{\label{fig:scheme_rades} Scheme of the five cavities coupled all-inductive RADES haloscope. The coaxial cable has been inserted in the first cavity.}
	\end{center}
\end{figure}

Finally, we have applied the present formulation to analyze the first all-inductive coupled five cavities haloscope designed by the RADES collaboration \cite{RADES_paper1}, which was successfully measured at the CERN Axion Solar Telescope (CAST) facility \cite{RADES_paper3} (see Figure~\ref{fig:scheme_rades}). The operation mode of each cavity is the fundamental $TE_{101}$ rectangular mode. The geometrical parameters of the filter can be found in \cite{RADES_paper1}. We have performed the simulations at cryogenic temperature ($2$~K) using this estimation for the electrical conductivity: $\sigma = 2 \cdot 10^9$ S/m. The resonant frequency used for axion detection is $f_1 \approx 8.4$~GHz. Again, a coaxial cable with characteristic impedance of $Z_0 = 50 \, \Omega$ ($b=0.635$~mm, $a=2.11$~mm, $\varepsilon_r=2.08$) has been inserted in the first cavity operating in critical coupling regime. The axion field and the axion-photon interaction model is described by the parameters  $g_{a \gamma \gamma} \, a_0 = -8.51 \cdot 10^{-22}$ \cite{kim_CAPP_2019}. The phase of the axion field used in the simulations is zero: $\varphi = 0$ rad. The external homogeneous static magnetic field used in the CAST experiment is oriented along the vertical direction: $\vec{B}_e = B_e \, \hat{y}$ with $B_e=8.8$~T. Thus, the axion current given in (\ref{axion_current_coaxial_single_mode_port}) becomes
\begin{eqnarray}      \label{I_axion_RADES}
	I_a \,  = \, \frac{1}{\mu_0}  \, g_{a \gamma \gamma} \, a_0 \, B_e \, j \, k \,
	\sum_{m=1}^{M}  \, \frac{\kappa_m}{\kappa_m^2- k^2}   \, \left( \int_{S(1)} \vec{H}_m(\vec{r})  \cdot \vec{h}_{TEM}(\vec{r}) \, dS \right) \,   \,  \left( \int_ V \vec{E}_m(\vec{r}) \cdot  \hat{y}  	\, dV   \right)   \nonumber  \\    
\end{eqnarray}

\begin{figure}[h!]
	\begin{center}   
		\begin{subfigure} [b]{0.8\textwidth}
	        \includegraphics[width= \textwidth]{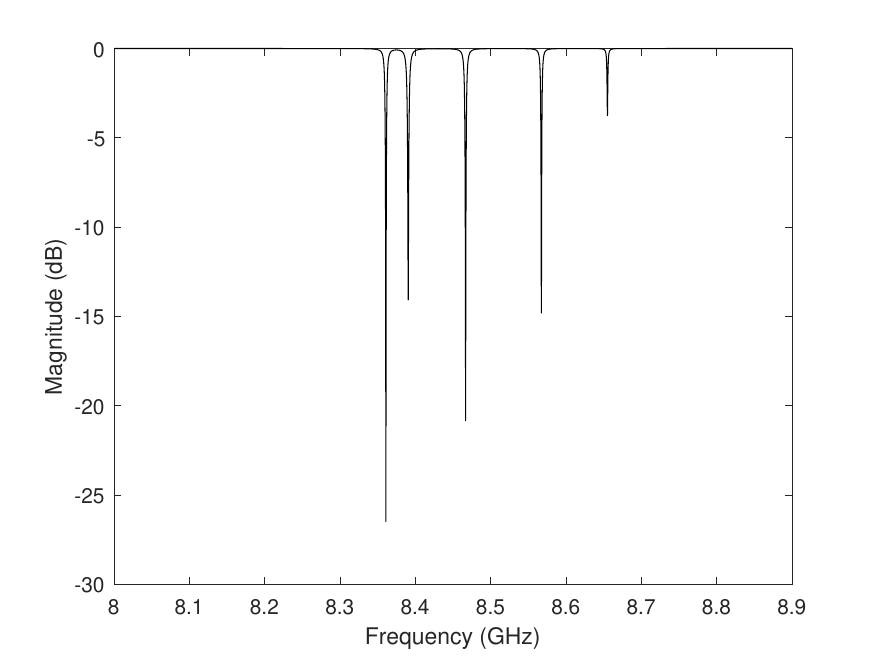}
	    \end{subfigure}
        \begin{subfigure} [b]{0.8\textwidth}
	        \includegraphics[width= \textwidth]{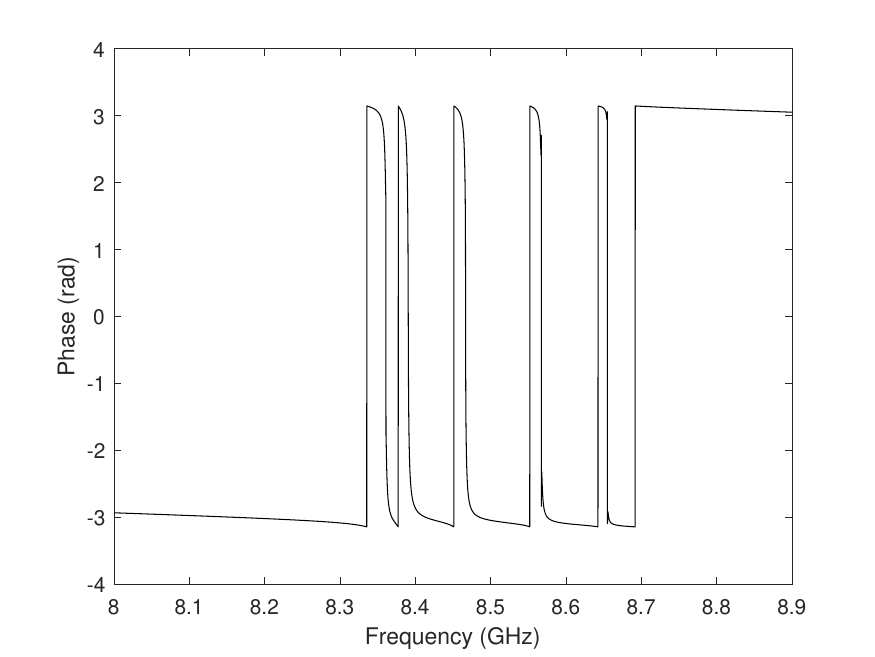}
	    \end{subfigure}	
		\caption{\label{fig:S11_rades} Reflection scattering parameter $S_{11}$ as a function of the frequency for the RADES haloscope. Magnitude (top) and phase (bottom) are shown.}
	\end{center}
\end{figure}

\begin{figure}[h!]
	\begin{center}   
		\begin{subfigure} [b]{0.8\textwidth}
	        \includegraphics[width= \textwidth]{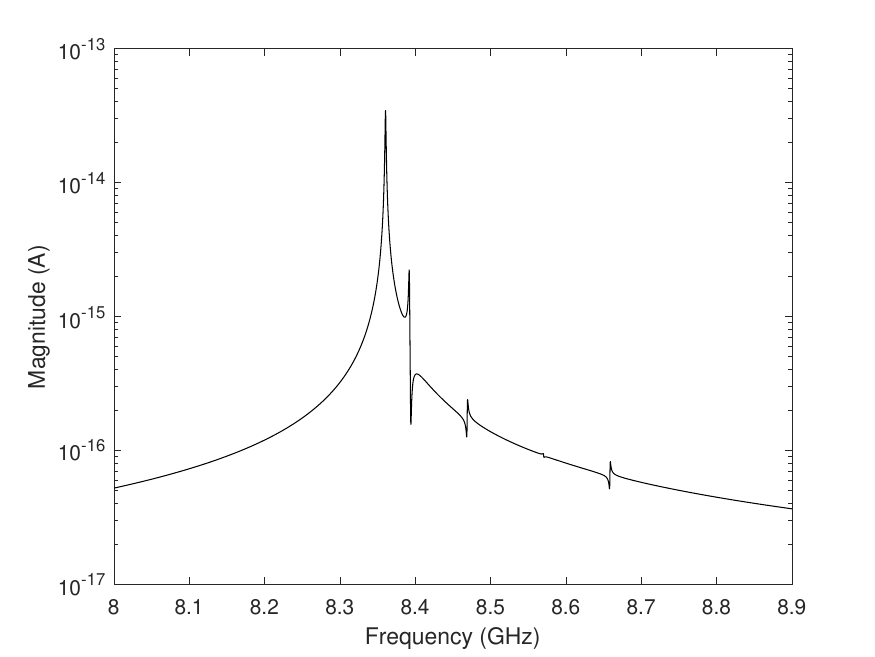}
	    \end{subfigure}
        \begin{subfigure} [b]{0.8\textwidth}
	        \includegraphics[width= \textwidth]{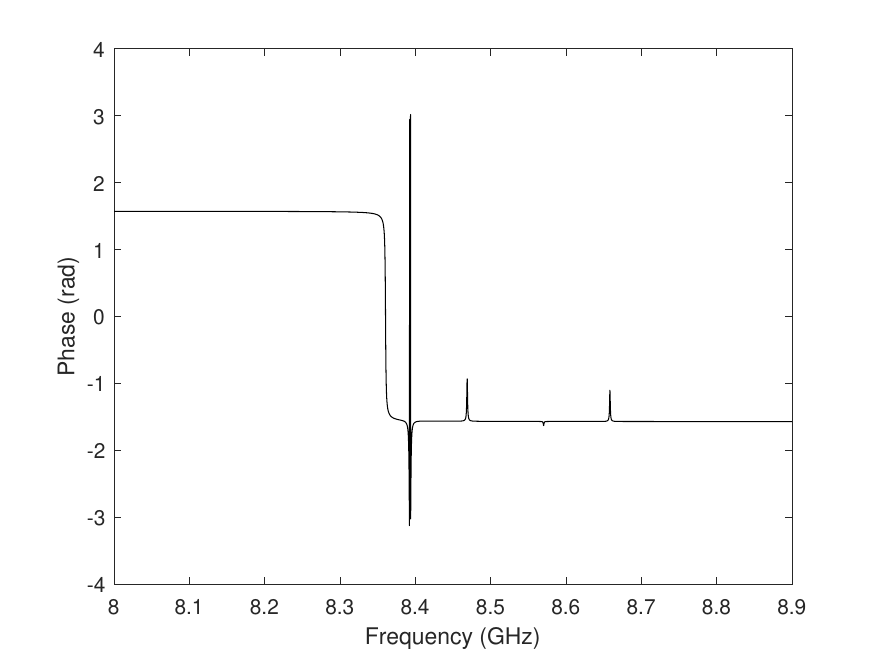}
	    \end{subfigure}
		\caption{\label{fig:Ia_rades} Axion current $I_a$ as a function of the frequency for the RADES haloscope, as obtained with the method presented in this paper. Magnitude (top) and phase (bottom) have been plotted.}
	\end{center}
\end{figure}

\begin{figure}[h!]
	\begin{center}   
		\begin{subfigure} [b]{0.8\textwidth}
	        \includegraphics[width= \textwidth]{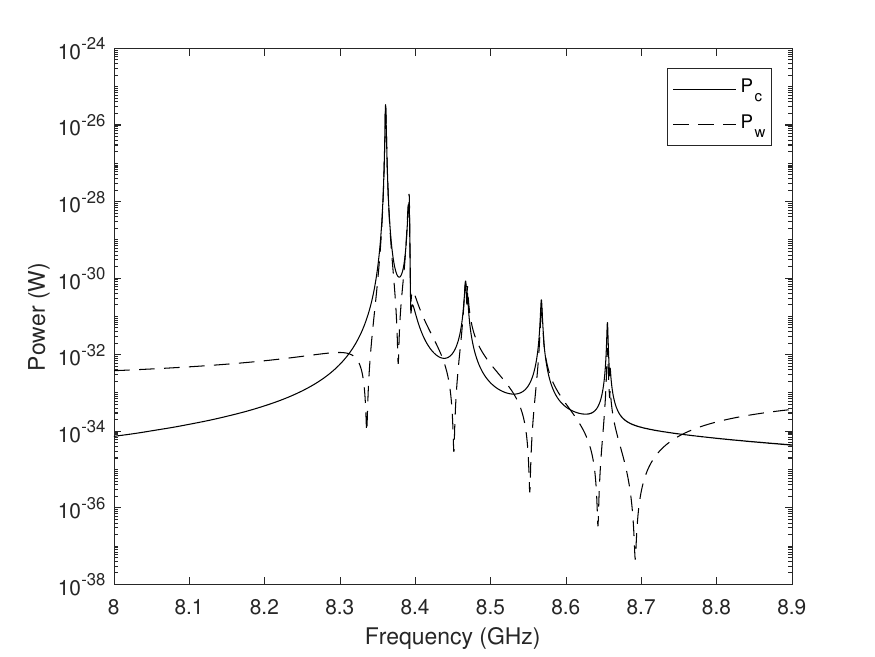}
	    \end{subfigure}
        \begin{subfigure} [b]{0.8\textwidth}
	        \includegraphics[width= \textwidth]{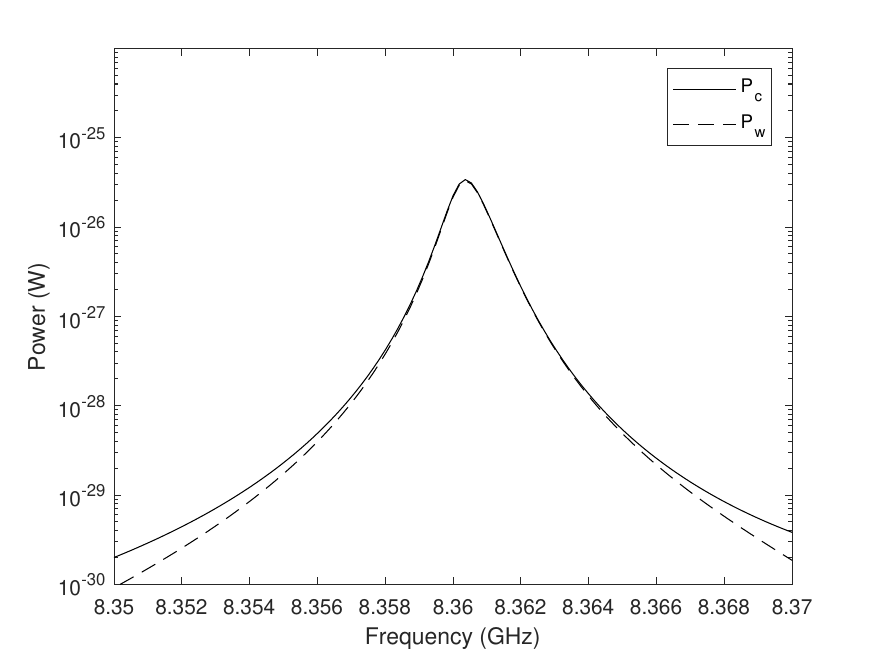}
	    \end{subfigure}
		\caption{\label{fig:power_rades} Extracted $P_w$ and dissipated $P_c$ powers as a function of the frequency for the RADES haloscope (top). In the bottom figure we present a zoom of the plot in order to check the coupling regime achieved in the design of the coaxial probe at the main resonance, observing a very accurate critical coupling condition.}
	\end{center}
\end{figure}

In Table~II we have summarized the most relevant parameters of the haloscope, including the electric field pattern presented in Figure~2 of \cite{RADES_paper1}.
\begin{table}
\begin{center}
\caption{Parameters of the RADES haloscope.}
	\begin{tabular}{|c|c|c|c|c|}    \hline
		$m$ & Eletric field pattern & $f_m$ (GHz) &  $Q_m$	& $C_m$  \\  \hline
		1     & $+ \, + \, + \, + \, +$  & $8.428$  &  $40386$  &  $0.65$  \\  \hline
		2     & $+ \, + \, 0 \, - \, -$  & $8.454$  &  $42033$  &  $3.2 \cdot 10^{-7}$  \\  \hline  
		3     & $- \, + \, + \, + \, -$  & $8.528$  &  $43654$  &  $8.1 \cdot 10^{-5}$  \\  \hline  
		4     & $- \, + \, 0 \, - \, +$  & $8.625$  &  $45882$  &  $1.6 \cdot 10^{-12}$  \\  \hline 
		5     & $- \, + \, - \, + \, -$  & $8.710$  &  $48048$  &  $6.4 \cdot 10^{-6}$  \\  \hline  
	\end{tabular}
\end{center}
\end{table}
In Figure~\ref{fig:S11_rades} we first plot the electrical response of the structure, observing the peaks of the five resonances. Second, we show in Figure~\ref{fig:Ia_rades} the magnitude and phase of the axion current (\ref{I_axion_RADES}), including the first $M=5$ resonant modes, which is maximum for the first resonance. The amplitude of the axion current is extremely low in the fourth resonance because the geometric form factor is negligible. Finally, in Figure~\ref{fig:power_rades} we have displayed both the extracted ($P_w$) and dissipated ($P_c$) powers, demonstrating that the critical coupling regime is not only satisfied at the first resonant peak (as designed) but also around the other resonances (see details for the first resonance in the zoom of Figure~\ref{fig:power_rades}).

\section{Conclusions}
In this paper, the well-known BI-RME 3D full-wave technique has been successfully adapted to the study of microwave haloscopes based on resonant cavities. The formulation has been derived from time-domain Maxwell's equations to account for the axion-photon interaction, and it considers the cavity wall losses of the resonating structure by means of the standard perturbation method. Then, classical Network Theory has been used to obtain the expressions for the current and voltage induced by the axion.

Following this technique, an exact frequency-domain expression for the current produced by the axion-photon coupling mechanism has been accurately obtained in complex form (magnitude and phase) for the first time, avoiding Cauchy-Lorentz approximations. The conditions for the electromagnetic analysis include microwave haloscopes based on resonant cavities of arbitrary geometries and number of access waveguide ports. A multimodal response for each port is obtained, i.e., if higher order modes are excited at the waveguide ports its response can be obtained for each higher order mode. In the present formulation the axion field is distributed within the whole cavity and the interaction is produced throughout all the volume. Since this method is able to calculate the coupling of the axion with each resonant mode, it allows to calculate the electromagnetic field pattern generated by the axion inside the cavity as a superposition of the different excited modes.

The derived general expression for the equivalent axion current takes into account the axion source interaction with both cavity and extracting probes. It shows how the electromagnetic energy generated by the axion-photon coupling is split into two terms: the extracted power (delivered to the ports) and the dissipated power (Ohmic losses). To the authors knowledge, available commercial full-wave electromagnetic simulators cannot deal nowadays with this axion-photon interaction problem.

In order to verify the proposed technique an accurate comparison with previous bibliography data has been carried out showing a good agreement nearby the main resonance peak. Additionally, precise results in a wide-band region around the working frequency have also been provided. This has allowed highlight the importance of higher order modes, showing that care must be taken since the whole experiment reliability can be affected by neighboring resonant modes. Depending on the nearness to the axion resonance and the form factor, the axion can be coupled to more than one mode. A well-designed experiment will try to separate enough the main mode from neighbouring modes in order to consider the modal overlapping negligible. 

Future research lines may include the consideration of dielectric and/or magnetic materials in the analysis. Also, multi-port configurations may be interesting when coherently combining the received signal in different coupled cavities for post-processing purposes, since the phase response of the photon-axion conversion is obtained. This last is a key feature of this technique: a complete information of the extracted signal (magnitude and phase) in a broad range of frequencies is obtained. This allows not only to calculate the extracted RF power, but also to study the phase of the axion field. One particularly promising possibility that can be explored consists in the analytical study of setups consisting on several cavities, placed at a distance for which the spatial gradient of the axion field might be non-negligible; opening a new door to directional sensitive experiments like the ones proposed in \cite{directionality_2018} .
In summary, it must be pointed out that the developed method can be accurately used for broad-band design purposes, when working with haloscopes based on resonant microwave cavities.

\paragraph{ACKOWLEDGEMENT}
This work has been funded by the Spanish Agencia Estatal de Investigacion (AEI) and Fondo Europeo de Desarrollo Regional (FEDER) under project PID2019-108122GB-C33, and supported by the grant FPI BES-2017-079787 (under project FPA-2016-76978-C3-2-P). JG  acknowledges  support  through  the  European Research Council under grant ERC-2018-StG-802836 (AxScale  project). 

\section{Appendix}
In this Appendix the most important properties of the Green's functions used in this work are summarized.
\subsection{Expansion of the electromagnetic field in a microwave cavity}
The electromagnetic energy stored in the volume $V$ of a cavity is finite, so the electric and magnetic fields of the cavity belong to the complex Hilbert space of square integrable functions $\mathcal{L}_2(V)$. Following the Helmholtz's theorem \cite{collin_FTGW}, the electromagnetic field existing within a cavity can be expanded at an infinite number of both solenoidal and irrotational modes \cite{collin_FMI}, \cite{collin_FTGW}, \cite{kurokawa}, \cite{conciauro}, \cite{vanbladel}. For a closed cavity we only need the solenoidal ones, but a cavity is usually coupled to the outside sources by means of a small aperture (access waveguide port) or a probe or loop connected to a coaxial transmision line, which also requires the inclusion of the irrotational modes. 

\subsubsection{Expansion of the electric field}
Taken this starting point, the expansion of the electric field eigenvectors is expressed in terms of solenoidal and irrotational modes:  \\
(a) Solenoidal modes: These modes satisfy the following differential equations:
\begin{eqnarray}
	\nabla^2 \vec{E}_i + k_i^2 \vec{E}_i & = & \vec{0} \mbox{ in } V \nonumber \\
	\nabla \cdot \vec{E}_i  =  0 \, \, & ; & \, \,  \nabla \times \vec{E}_i \ne  \vec{0} \mbox{ in } V \nonumber \\
	\vec{n} \times \vec{E}_i & = & \vec{0} \mbox{ on } S_V \nonumber
\end{eqnarray}
where $S_V$ is the surface of the cavity, and $k_i^2$ and $\vec{E}_i$ are the solenoidal electric eigenvalues and eigenvectors, respectively. These set of modes correspond to the physical resonances.   \\
(b) Irrotational modes: These modes satisfy the following differential equations:
\begin{eqnarray}
	\nabla^2 \vec{f}_i + \mu_i^2 \vec{f}_i & = & \vec{0} \mbox{ in } V \\
	\nabla \times \vec{f}_i =  \vec{0} \,  & ; &  \, \nabla \cdot \vec{f}_i \ne  0 \mbox{ in } V \\
	\vec{n} \times \vec{f}_i  & = & \vec{0} \mbox{ on } S_V \nonumber  \\
	\mu_i \, \vec{f}_i & = & \nabla v_i  \\
	\nabla^2 v_i + \mu_i^2 v_i  & = & 0 \mbox{ in } V  \\
	v_i & = & 0 \mbox{ on } S_V \mbox{ (Dirichlet boundary condition) }
\end{eqnarray}
where $\mu_i^2$ and $\vec{f}_i$ are the irrotational electric eigenvalues and eigenvectors, respectively. The scalar eigenfunctions $v_i$ are used to obtain the corresponding eigenvectors.

\subsubsection{Expansion of the magnetic field}
The expansion of the magnetic field eigenvectors is also expressed in terms of solenoidal and irrotational modes:  \\
(a) Solenoidal modes: These modes satisfy the following differential equations:
\begin{eqnarray}
	\nabla^2 \vec{H}_i + k_i^2 \vec{H}_i  & = & \vec{0} \mbox{ in } V \nonumber  \\
	\nabla \cdot \vec{H}_i =  0  \, \, & ; & \, \,  \nabla \times \vec{H}_i \ne  \vec{0}  \mbox{ in } V \nonumber \\
	\vec{n} \cdot \vec{H}_i & = & 0 \mbox{ on } S_V \nonumber
\end{eqnarray}
where $k_i^2$ and $\vec{H}_i$ are the solenoidal magnetic eigenvalues and eigenvectors, respectively. These set of modes correspond to the physical resonances, and are related with the solenoidal electric modes as follows:
\begin{eqnarray}
	\nabla \times \vec{E}_i = k_i \vec{H}_i
	\, \, \, \, \, , \, \, \, \, \, \nabla \times \vec{H}_i =
	k_i\vec{E}_i  \nonumber
\end{eqnarray}
(b) Irrotational modes: These modes satisfy the following differential equations:
\begin{eqnarray}
	\nabla^2 \vec{g}_i + \nu_i^2 \vec{g}_i & = & \vec{0} \mbox{ in } V \nonumber \\
	\nabla \times \vec{g}_i  =  \vec{0} \,  & ; &  \, \nabla \cdot \vec{g}_i \ne  0 \mbox{ in } V \nonumber \\
	\vec{n} \cdot \vec{g}_i  & = &  0 \mbox{ on } S_V \nonumber  \\
	\nu_i \, \vec{g}_i & = & \nabla w_i  \\
	\nabla^2 w_i + \nu_i^2 w_i  & =  & 0 \mbox{ in } V  \\
	\frac{\partial \, w_i}{\partial \, n} & = & 0 \mbox{ on } S_V \mbox{ (Neumann boundary condition) }
\end{eqnarray}
where $\nu_i^2$ and $\vec{g}_i$ are the irrotational magnetic eigenvalues and eigenvectors, respectively. The scalar eigenfunctions $w_i$ are used to obtain the corresponding eigenvectors.

\subsubsection{Orthonormalization properties and modal expansion}
The previous modes satisfy these orthonormalization relationships:
\begin{eqnarray}
	\int_V \vec{E}_i \cdot \vec{E}_j \, dV \, = \, \delta_{i,j}
	\, \, \, \, \, ; \, \, \, \, \, \int_V \vec{f}_i \cdot \vec{f}_j
	\, dV \, = \, \delta_{i,j}  \, \, \, \, \, ; \, \, \, \, \, \int_V \vec{E}_i \cdot \vec{f}_j \, dV \, = \, 0  \nonumber \\
	\int_V \vec{H}_i \cdot \vec{H}_j \, dV \, = \, \delta_{i,j} \, \, \, \, \, ;
	\, \, \, \, \, \int_V \vec{g}_i \cdot \vec{g}_j
	\, dV \, = \, \delta_{i,j}  \, \, \, \, \, ; \, \, \, \, \, \int_V \vec{H}_i \cdot \vec{g}_j \, dV \, = \, 0 \, . \nonumber
\end{eqnarray}
where $\delta_{i,j}$ is the Kronecker delta. These properties allow to obtain the modal expansion coefficients as
\begin{eqnarray}
	E_i \, = \, 	\int_V \vec{E} \cdot \vec{E}_i \, dV 
	\, \, \, \, \, ; \, \, \, \, \, F_i \, = \, \int_V \vec{E} \cdot \vec{f}_i
	\, dV  \nonumber \\
	H_i \, = \,	\int_V \vec{H} \cdot \vec{H}_i \, dV  \, \, \, \, \, ;
	\, \, \, \, \, G_i \, = \, \int_V \vec{H} \cdot \vec{g}_i  
	\, dV \, . \nonumber
\end{eqnarray}
Finally we can develop the modal expansion for the electromagnetic field within the resonator, as
\begin{eqnarray}
	\vec{E} & = & \sum_i E_i \vec{E}_i + \sum_i F_i \vec{f}_i   \nonumber  \\
	\vec{H} & = & \sum_i H_i \vec{H}_i + \sum_i G_i \vec{g}_i  \,  . \nonumber
\end{eqnarray}

\subsection{Potential Green's functions}
BI-RME 3D formalism is based on the use of the electric and magnetic scalar and dyadic potentials under the Coulomb gauge defined on a cavity \cite{birme3d_overview}, \cite{conciauro}. We want to emphasize that both  electric scalar $g^e$ and dyadic $\mathbf{\vec{G}^{\rm A}}$ potentials are related with the real electric charge and currents densities existing on the conducting walls of the cavity. However, the magnetic scalar $g^m$ and dyadic $\mathbf{\vec{G}^{\rm F}}$ potentials represent fictitious magnetic charge and current densities that allow to introduce in the formulation the existence of the access waveguide ports on the surface cavity. These fictitious magnetic charges and currents are used to impose the correct boundary conditions for the tangential components of the electromagnetic fields across the port interfaces.
We want to remark that fictitious magnetic charges and currents have been typically used in Classical Electromagnetic Theory in waveguides, cavities and antennas problems \cite{harrington}, where they are not related with magnetic monopoles \cite{jackson}.

\subsubsection{Electric potential Green´s functions}
The electric scalar static potential $g^e$ satisfies
\begin{eqnarray}
	\nabla^2 g^e(\vec{r},\vec{r'}) & = & - \delta(\vec{r} - \vec{r'}) \mbox{ in } V  \nonumber \\
	g^e(\vec{r},\vec{r'}) & = & 0 \mbox{ on } S_V \nonumber
\end{eqnarray}
whose solution under the Columb gauge is
\begin{eqnarray}
	g^e(\vec{r},\vec{r'}) =  \sum_{i=1}^{+ \infty} \frac{1}{\mu_i^2} \,
	v_i(\vec{r}) v_i(\vec{r'}) \, .  \nonumber
\end{eqnarray}
This scalar Green's function can be split into a singular $g^e_s$ and a regular $g^e_r$ parts as follows:
\begin{eqnarray}
	g^e(\vec{r},\vec{r'}) & = & g^e_s(\vec{r},\vec{r'}) + g^e_r(\vec{r},\vec{r'}) \nonumber \\
	g^e_s(\vec{r},\vec{r'}) & = & \frac{1}{4 \pi R} \nonumber
\end{eqnarray}
where we have included the explicit expression of the singular part, which is the scalar potential Green's function in free space because the singularity does not depend on the boundary conditions; $R$ is the magnitude of the relative vector $\vec{R} \equiv \vec{r} - \vec{r'}$, $R \equiv ||\vec{R}||$.

The electric dyadic potential $\mathbf{\vec{G}^{\rm A}}$ satisfies
\begin{eqnarray}
	\nabla \times \nabla \times \mathbf{\vec{G}^{\rm
			A}}(\vec{r},\vec{r'}) - k^2 \mathbf{\vec{G}^{\rm
			A}}(\vec{r},\vec{r'}) & = &  \mathbf{\vec{I}} \delta(\vec{r} - \vec{r'})
	- \nabla \, \nabla' g^E(\vec{r},\vec{r'})  \mbox{ in } V \nonumber  \\
	\vec{n} \times \mathbf{\vec{G}^{\rm A}} & = & 0 \mbox{ on } S_V
	\nonumber
\end{eqnarray}
whose solution under the Columb gauge is expressed as a mode expansion:
\begin{eqnarray}
	\mathbf{\vec{G}^{\rm A}}(\vec{r},\vec{r'}) = \sum_{i=1}^{+ \infty} \,
	\frac{\vec{E}_i(\vec{r}) \, \vec{E}_i(\vec{r'})}{k_i^2 - k^2}  \, . \nonumber 
\end{eqnarray}
The convergence of this eigenvector series can be accelerated by extracting its zero-frequency limit because the singularity is independent of the frequency. Thus, the electric dyadic potential can be divided in two terms: the static part $\mathbf{\vec{G}_0^{\rm A}}$ which is singular and does not depend on the frequency, and the frequency dependent term $\mathbf{\vec{G}_r^{\rm A}}$. In this scenario, it is possible to demonstrate that the static part can be expressed as the summation of the singular term and the regular static term $\mathbf{\vec{G}_{0_r}^{\rm A}}$. Finally we obtain
\begin{eqnarray}
	\mathbf{\vec{G}^{\rm A}}(\vec{r},\vec{r'}) & = & 
	\mathbf{\vec{G}_0^{\rm
			A}}(\vec{r},\vec{r'}) + \mathbf{\vec{G}_r^{\rm A}}(\vec{r},\vec{r'}) \nonumber \\
	\mathbf{\vec{G}_0^{\rm
			A}}(\vec{r},\vec{r'}) & = & \; \frac{1}{8 \pi R} \; \left(
	\mathbf{\vec{I}} \; + \; \frac{\vec{R} \, \vec{R}}{R^2} \right) + \mathbf{\vec{G}_{0_r}^{\rm A}}(\vec{r},\vec{r'}) \nonumber \\
	\mathbf{\vec{G}_r^{\rm A}}(\vec{r},\vec{r'}) & = & k^2 \; \sum_{i=1}^{+ \infty}
	\; \frac{\vec{E}_i(\vec{r}) \, \vec{E}_i(\vec{r'})}{k_i^2 (k_i^2 -
		k^2)}  \nonumber
\end{eqnarray}
The symbol $\mathbf{\vec{I}}$ represents the unit dyadic.

\subsubsection{Magnetic potential Green´s functions}
The magnetic potential Green's functions are very similar to the electric potential Green's functions. We summarize the most important equations, using duality on the previous equations:
\begin{eqnarray}
	\nabla^2 g^m(\vec{r},\vec{r'}) & = & - \delta(\vec{r} - \vec{r'}) \mbox{ in } V   \nonumber \\
	\frac{\partial g^m}{\partial n} & = & 0 \mbox{ on } S_V \nonumber
\end{eqnarray}
\begin{eqnarray}
	g^m(\vec{r},\vec{r'}) & = & \sum_{i=1}^{+ \infty} \frac{1}{\nu_i^2} \,
	w_i(\vec{r}) w_i(\vec{r'}) =
	g^m_s(\vec{r},\vec{r'}) + g^m_r(\vec{r},\vec{r'}) \nonumber \\
	g^m_s(\vec{r},\vec{r'}) & = & \frac{1}{4 \pi R}  \nonumber
\end{eqnarray}
\begin{eqnarray}
	\nabla \times \nabla \times \mathbf{\vec{G}^{\rm
			F}}(\vec{r},\vec{r'}) - k^2 \mathbf{\vec{G}^{\rm
			F}}(\vec{r},\vec{r'}) & = &  \mathbf{\vec{I}} \delta(\vec{r} -
	\vec{r'})
	- \nabla \, \nabla' g^m(\vec{r},\vec{r'})  \mbox{ in } V   \nonumber \\
	\vec{n} \times \mathbf{\nabla \times \vec{G}^{\rm F}} & = & 0
	\mbox{ on } S_V \nonumber
\end{eqnarray}
\begin{eqnarray}
	\mathbf{\vec{G}^{\rm F}}(\vec{r},\vec{r'}) & = & \sum_{i=1}^{+ \infty} \,
	\frac{\vec{H}_i(\vec{r}) \, \vec{H}_i(\vec{r'})}{k_i^2 - k^2} =
	\mathbf{\vec{G}_0^{\rm
			F}}(\vec{r},\vec{r'}) + \mathbf{\vec{G}_r^{\rm F}}(\vec{r},\vec{r'})  \nonumber \\
	\mathbf{\vec{G}_0^{\rm
			F}}(\vec{r},\vec{r'}) & = & \; \frac{1}{8 \pi R} \; \left(
	\mathbf{\vec{I}} \; + \; \frac{\vec{R} \, \vec{R}}{R^2} \right) + \mathbf{\vec{G}_{0_r}^{\rm F}}(\vec{r},\vec{r'}) \nonumber \\
	\mathbf{\vec{G}_r^{\rm F}}(\vec{r},\vec{r'}) & = & k^2 \; \sum_{i=1}^{+ \infty}
	\; \frac{\vec{H}_i(\vec{r}) \, \vec{H}_i(\vec{r'})}{k_i^2 (k_i^2 -
		k^2)}  \nonumber
\end{eqnarray}
%

\bibliography{mybibfile.bib}

\begin{thebibliography}{10}
\expandafter\ifx\csname url\endcsname\relax
  \def\url#1{\texttt{#1}}\fi
\expandafter\ifx\csname urlprefix\endcsname\relax\def\urlprefix{URL }\fi
\expandafter\ifx\csname href\endcsname\relax
  \def\href#1#2{#2} \def\path#1{#1}\fi

\bibitem{Sikivie:1983ip}
P.~Sikivie, {Experimental Tests of the Invisible Axion}, Phys. Rev. Lett. 51
  (1983) 1415--1417, [Erratum: Phys.Rev.Lett. 52, 695 (1984)].
\newblock \href {http://dx.doi.org/10.1103/PhysRevLett.51.1415}
  {\path{doi:10.1103/PhysRevLett.51.1415}}.

\bibitem{Sikivie:2021}
P.~Sikivie, {Invisible axion search methods}, Rev. Modern Physics 93 (2021)
  015004.
\newblock \href {http://dx.doi.org/0034-6861=2021=93(1)=015004(36)}
  {\path{doi:0034-6861=2021=93(1)=015004(36)}}.

\bibitem{Asztalos:2001tf}
S.~J. Asztalos, et~al., {Large scale microwave cavity search for dark matter
  axions}, Phys. Rev. D 64 (2001) 092003.
\newblock \href {http://dx.doi.org/10.1103/PhysRevD.64.092003}
  {\path{doi:10.1103/PhysRevD.64.092003}}.

\bibitem{Braine:2019fqb}
T.~Braine, et~al., {Extended Search for the Invisible Axion with the Axion Dark
  Matter Experiment}, Phys. Rev. Lett. 124~(10) (2020) 101303.
\newblock \href {http://arxiv.org/abs/1910.08638} {\path{arXiv:1910.08638}},
  \href {http://dx.doi.org/10.1103/PhysRevLett.124.101303}
  {\path{doi:10.1103/PhysRevLett.124.101303}}.

\bibitem{Alesini:2017ifp}
D.~Alesini, D.~Babusci, D.~Di~Gioacchino, C.~Gatti, G.~Lamanna, C.~Ligi, {The
  KLASH Proposal}\href {http://arxiv.org/abs/1707.06010}
  {\path{arXiv:1707.06010}}.

\bibitem{Kenany:2016tta}
S.~Al~Kenany, et~al., {Design and operational experience of a microwave cavity
  axion detector for the 20\textendash{}100 $\mu$eV range}, Nucl. Instrum.
  Meth. A 854 (2017) 11--24.
\newblock \href {http://arxiv.org/abs/1611.07123} {\path{arXiv:1611.07123}},
  \href {http://dx.doi.org/10.1016/j.nima.2017.02.012}
  {\path{doi:10.1016/j.nima.2017.02.012}}.

\bibitem{McAllister:2017lkb}
B.~T. McAllister, G.~Flower, E.~N. Ivanov, M.~Goryachev, J.~Bourhill, M.~E.
  Tobar, {The ORGAN Experiment: An axion haloscope above 15 GHz}, Phys. Dark
  Univ. 18 (2017) 67--72.
\newblock \href {http://arxiv.org/abs/1706.00209} {\path{arXiv:1706.00209}},
  \href {http://dx.doi.org/10.1016/j.dark.2017.09.010}
  {\path{doi:10.1016/j.dark.2017.09.010}}.

\bibitem{Alesini:2020vny}
D.~Alesini, et~al., {Search for invisible axion dark matter of mass
  m$_a=43~\mu$eV with the QUAX--$a\gamma$ experiment}, Phys. Rev. D 103~(10)
  (2021) 102004.
\newblock \href {http://arxiv.org/abs/2012.09498} {\path{arXiv:2012.09498}},
  \href {http://dx.doi.org/10.1103/PhysRevD.103.102004}
  {\path{doi:10.1103/PhysRevD.103.102004}}.

\bibitem{Jeong:2017hqs}
J.~Jeong, S.~Youn, S.~Ahn, J.~E. Kim, Y.~K. Semertzidis, {Concept of
  multiple-cell cavity for axion dark matter search}, Phys. Lett. B 777 (2018)
  412--419.
\newblock \href {http://arxiv.org/abs/1710.06969} {\path{arXiv:1710.06969}},
  \href {http://dx.doi.org/10.1016/j.physletb.2017.12.066}
  {\path{doi:10.1016/j.physletb.2017.12.066}}.

\bibitem{RADES_paper1}
A.~Álvarez Melcón, S.~Arguedas-Cuendis, C.~Cogollos, A.~Díaz-Morcillo,
  B.~Döbrich, J.~D. Gallego, B.~Gimeno, I.~G. Irastorza, A.~J.
  Lozano-Guerrero, C.~Malbrunot, P.~Navarro, C.~Peña-Garay, J.~Redondo,
  T.~Vafeiadis, W.~Wuensch, Axion searches with microwave filters: the rades
  project, Journal of Cosmology and Astroparticle Physics 040 (2018) 1 -- 22.
\newblock \href {http://dx.doi.org/10.1088/1475-7516/2018/05/040}
  {\path{doi:10.1088/1475-7516/2018/05/040}}.

\bibitem{TheMADMAXWorkingGroup:2016hpc}
A.~Caldwell, G.~Dvali, B.~Majorovits, A.~Millar, G.~Raffelt, J.~Redondo,
  O.~Reimann, F.~Simon, F.~Steffen, {Dielectric Haloscopes: A New Way to Detect
  Axion Dark Matter}, Phys. Rev. Lett. 118~(9) (2017) 091801.
\newblock \href {http://arxiv.org/abs/1611.05865} {\path{arXiv:1611.05865}},
  \href {http://dx.doi.org/10.1103/PhysRevLett.118.091801}
  {\path{doi:10.1103/PhysRevLett.118.091801}}.

\bibitem{Irastorza:2018dyq}
I.~G. Irastorza, J.~Redondo, {New experimental approaches in the search for
  axion-like particles}, Prog. Part. Nucl. Phys. 102 (2018) 89--159.
\newblock \href {http://arxiv.org/abs/1801.08127} {\path{arXiv:1801.08127}},
  \href {http://dx.doi.org/10.1016/j.ppnp.2018.05.003}
  {\path{doi:10.1016/j.ppnp.2018.05.003}}.

\bibitem{Peng:2000hd}
H.~Peng, et~al., {Cryogenic cavity detector for a large scale cold dark-matter
  axion search}, Nucl. Instrum. Meth. A 444 (2000) 569--583.
\newblock \href {http://dx.doi.org/10.1016/S0168-9002(99)00971-7}
  {\path{doi:10.1016/S0168-9002(99)00971-7}}.

\bibitem{Baker:2011na}
O.~K. Baker, M.~Betz, F.~Caspers, J.~Jaeckel, A.~Lindner, A.~Ringwald,
  Y.~Semertzidis, P.~Sikivie, K.~Zioutas, {Prospects for Searching Axion-like
  Particle Dark Matter with Dipole, Toroidal and Wiggler Magnets}, Phys. Rev. D
  85 (2012) 035018.
\newblock \href {http://arxiv.org/abs/1110.2180} {\path{arXiv:1110.2180}},
  \href {http://dx.doi.org/10.1103/PhysRevD.85.035018}
  {\path{doi:10.1103/PhysRevD.85.035018}}.

\bibitem{RADES_paper2}
A.~Álvarez Melcón, S.~Arguedas-Cuendis, C.~Cogollos, A.~Díaz-Morcillo,
  B.~Döbrich, J.~D. Gallego, J.~M. García-Barceló, B.~Gimeno, J.~Golm, I.~G.
  Irastorza, A.~J. Lozano-Guerrero, C.~Malbrunot, A.~Millar, P.~Navarro,
  C.~Peña-Garay, J.~Redondo, W.~Wuensch, Scalable haloscopes for axion dark
  matter detection in the 30 microev range with rades, Journal of High Energy
  Physics 084 (2020) 1 -- 28.
\newblock \href {http://dx.doi.org/10.1007/JHEP07(2020)084}
  {\path{doi:10.1007/JHEP07(2020)084}}.

\bibitem{kim_CAPP_2019}
Y.~Kim, D.~Kim, J.~Jeong, J.~Kim, Y.~C. Shin, Y.~K. Semertzidis, Effective
  approximation of electromagnetism for axion haloscope searches, Physics of
  the Dark Universe 26~(100362) (2019) 1 -- 10.
\newblock \href {http://dx.doi.org/10.1016/j.dark.2019.100362}
  {\path{doi:10.1016/j.dark.2019.100362}}.

\bibitem{birme3d_3Dcavities}
P.~Arcioni, M.~Bressan, L.~Perregrini, A new boundary integral approach to the
  determination of resonant modes of arbitrarily shaped cavities, IEEE
  Transactions on Microwave Theory and Techniques MTT-43~(8) (1995) 1848 --
  1856.
\newblock \href {http://dx.doi.org/10.1109/22.402270}
  {\path{doi:10.1109/22.402270}}.

\bibitem{birme3d_3Dcavities_ports}
P.~Arcioni, M.~Bozzi, M.~Bressan, G.~Conciauro, L.~Perregrini,
  Frequency/time-domain modelling of 3d waveguide structures by a bi-rme
  approach, International Journal of Numerical Modeling 15 (2002) 3 -- 21.
\newblock \href {http://dx.doi.org/10.1002/jnm.429}
  {\path{doi:10.1002/jnm.429}}.

\bibitem{birme3d_fermin}
F.~Mira, M.~Bressan, G.~Conciauro, B.~Gimeno, V.~Boria, Fast s-domain modeling
  of rectangular waveguides with radially symmetric metal insets, IEEE
  Transactions on Microwave Theory and Techniques 53~(4) (2005) 1294 -- 1303.
\newblock \href {http://dx.doi.org/10.1109/TMTT.2005.845762}
  {\path{doi:10.1109/TMTT.2005.845762}}.

\bibitem{birme3d_angel_posts}
Ángel A.~San~Blas, F.~Mira, V.~E. Boria, B.~Gimeno, M.~Bressan, P.~Arcioni, On
  the fast and rigorous analysis of compensated waveguide junctions using
  off-centered partial-height metallic posts, IEEE Transactions on Microwave
  Theory and Techniques 55~(1) (2007) 168 -- 175.
\newblock \href {http://dx.doi.org/10.1109/TMTT.2006.886928}
  {\path{doi:10.1109/TMTT.2006.886928}}.

\bibitem{birme3d_jordi_MWCL}
J.~Gil, A.~M. Pérez, B.~Gimeno, M.~Bressan, V.~Boria, G.~Conciauro, Analysis
  of cylindrical dielectric resonators in rectangular cavities using a
  state-space integral-equation method, IEEE Microwave and Wireless Components
  Letters 16~(12) (2006) 636 -- 638.
\newblock \href {http://dx.doi.org/10.1109/LMWC.2006.885584}
  {\path{doi:10.1109/LMWC.2006.885584}}.

\bibitem{birme3d_jordi_MTT}
J.~Gil, A.~S. Blas, C.~Vicente, B.~Gimeno, M.~Bressan, V.~Boria, G.~Conciauro,
  M.~Maestre, Full-wave analysis and design of dielectric-loaded waveguide
  filters using a state-space integral-equation method, IEEE Transactions on
  Microwave Theory and Techniques 57~(1) (2009) 109 -- 120.
\newblock \href {http://dx.doi.org/10.1109/TMTT.2008.2008974}
  {\path{doi:10.1109/TMTT.2008.2008974}}.

\bibitem{birme3d_pavia_MTT}
M.~Bressan, S.~Battistutta, M.~Bozzi, L.~Perregrini, Modeling of inhomogeneous
  and lossy waveguide components by the segmentation technique combined with
  the calculation of green’s function by ewald’s method, IEEE Transactions
  on Microwave Theory and Techniques 66~(2) (2018) 633 -- 642.
\newblock \href {http://dx.doi.org/10.1109/TMTT.2017.2787587}
  {\path{doi:10.1109/TMTT.2017.2787587}}.

\bibitem{birme3d_overview}
P.~Arcioni, M.~Bozzi, M.~Bressan, G.~Conciauro, L.~Perregrini, The bi-rme
  method: an historical overview, 2014 International Conference on Numerical
  Electromagnetic Modeling and Optimization for RF, Microwave, and Terahertz
  Applications (NEMO)\href {http://dx.doi.org/10.1109/NEMO.2014.6995653}
  {\path{doi:10.1109/NEMO.2014.6995653}}.

\bibitem{conciauro}
G.~Conciauro, M.~Guglielimi, R.~Sorrentino, Advanced Modal Analysis. CAD
  Techniques for Waveguide Components and Filters, 1st Edition, John Wiley and
  Sons, Ltd, 2000.

\bibitem{chentotai}
C.-T. Tai, Generalized Vector and Dyadic Analysis. Applied Mathematics in Field
  Theory, 1st Edition, IEEE Press, 1991.

\bibitem{marcuvitz}
N.~Marcuvitz, Waveguide Handbook, 1st Edition, Peter Peregrinus Ltd, 1993.

\bibitem{hanson_yakovlev}
G.~W. Hanson, A.~B. Yakovlev, Operator Theory for Electromagnetics. An
  introduction, 1st Edition, Springer-Verlag, 2002.

\bibitem{collin_FMI}
R.~E. Collin, Foundations for Microwave Engineering, 2nd Edition, McGraw-Hill,
  Inc., 1992.

\bibitem{pozar}
D.~M. Pozar, Microwave Engineering, 4th Edition, John Wiley and Sons, Inc.,
  2012.

\bibitem{birme3d_angel_multipactor}
A.~San-Blas, B.~Gimeno, V.~Boria, Study of the multipactor phenomenon using a
  full-wave integral equation technique, International Journal of Electronics
  and Communications 79 (2017) 286 -- 290.
\newblock \href {http://dx.doi.org/10.1016/j.aeue.2017.06.009}
  {\path{doi:10.1016/j.aeue.2017.06.009}}.

\bibitem{jackson}
J.~D. Jackson, Classical Electrodynamics, 3rd Edition, John Wiley and Sons,
  Inc., 1999.

\bibitem{collin_FTGW}
R.~E. Collin, Field Theory of Guided Waves, 2nd Edition, IEEE Press, 1991.

\bibitem{kurokawa}
K.~Kurokawa, An Introduction to the Theory of Microwave Circuits, 1st Edition,
  Academic Press, 1969.

\bibitem{kim_CAPP_2020}
D.~Kim, J.~Jeong, S.~Youn, Y.~Kima, Y.~K. Semertzidis, Revisiting the detection
  rate for axion haloscopes, Journal of Cosmology and Astroparticle Physics
  2020~(03) (2020) 1 -- 14.
\newblock \href {http://dx.doi.org/10.1088/1475-7516/2020/03/066}
  {\path{doi:10.1088/1475-7516/2020/03/066}}.

\bibitem{CST}
\uppercase{CST STUDIO SUITE}: Electromagnetic field simulation software,
  \url{https://www.3ds.com/products-services/simulia/products/cst-studio-suite/}.

\bibitem{MATLAB}
\uppercase{MATLAB,} \url{https://mathworks.com/products/matlab.html}.

\bibitem{RADES_paper3}
A.~Álvarez Melcón, S.~Arguedas-Cuendis, J.~Baier, K.~Barth, H.~Bräuninger,
  S.~Calatroni, G.~Cantatore, F.~Caspers, J.~Castel, S.~Cetin, C.~Cogollos,
  T.~Dafni, M.~Davenport, A.~Dermenev, K.~Desch, A.~Díaz-Morcillo,
  B.~Döbrich, H.~Fischer, W.~Funk, J.~Gallego, J.~García-Barceló,
  A.~Gardikiotis, J.~Garza, B.~Gimeno, S.~Gninenko, J.~Golm, M.~H.~D. Hoffmann,
  I.~Irastorza, K.~Jakovcic, J.~Kaminski, M.~Karuza, B.~Lakic, J.~Laurent,
  A.~Lozano-Guerrero, G.~Luzón, C.~Malbrunot, M.~Maroudas, J.~Miralda,
  H.~Mirallas, L.~Miceli, P.~Navarro, A.~Ozbey, K.~Ozbozduman, C.~Peña-Garay,
  M.~Pivovaro, J.~Redondo, J.~Ruz, E.~R. Chóliz, S.~Schmidt, M.~Schumann,
  Y.~Semertzidis, S.~Solanki, L.~Stewart, I.~Tsagris, T.~Vafeiadis, J.~Vogel,
  W.~Wuensch, K.~Zioutas, First results of the cast-rades haloscope search for
  axions at 34.67 microev, \url{https://arxiv.org/abs/2104.13798}.

\bibitem{directionality_2018}
{S. Knirck, A. J. Millar, C. A. J. O’Hare, J. Redondo, F. D. Steffen},
  {Directional axion detection}, {Journal of Cosmology and Astroparticle
  Physics} 11 (2018) 051.
\newblock \href {http://dx.doi.org/10.1088/1475-7516/2018/11/051}
  {\path{doi:10.1088/1475-7516/2018/11/051}}.

\bibitem{vanbladel}
J.~V. Bladel, Electromagnetic Fields, 2nd Edition, IEEE Press, 2007.

\bibitem{harrington}
R.~F. Harrington, Time-Harmonic Electromagnetic Fields, 1st Edition, IEEE
  Press, 2001.

\end{thebibliography}
\end{document}